\documentclass[preprint, 11pt,3p,authoryear]{elsarticle}

\biboptions{longnamesfirst}

\usepackage[utf8]{inputenc}
\usepackage[T1]{fontenc}
\usepackage{lmodern}
\usepackage{microtype}
\usepackage{amsmath,amssymb,amsthm}
\usepackage{graphicx,booktabs,threeparttable,makecell,multirow,tabularx,siunitx}
\usepackage[table]{xcolor}
\usepackage{tikz}
\usetikzlibrary{arrows.meta,positioning,calc}
\usepackage{caption,enumitem,forest}
\usepackage[colorlinks=true,allcolors=blue]{hyperref}
\usepackage[nameinlink,noabbrev]{cleveref}
\usepackage{algorithm,algpseudocode}
\algrenewcommand\algorithmicrequire{\emph{Input:}}
\algrenewcommand\algorithmicensure{\emph{Output:}}

\newcommand{\E}{\mathbb{E}}
\newcommand{\TPR}{\mathrm{TPR}}
\newcommand{\FPR}{\mathrm{FPR}}
\newcommand{\TNR}{\mathrm{TNR}}
\newcommand{\TP}{\mathrm{TP}}
\newcommand{\TN}{\mathrm{TN}}
\newcommand{\FP}{\mathrm{FP}}
\newcommand{\FN}{\mathrm{FN}}

\begin{document}
    \begin{frontmatter}

    \title{{An Imbalance-Robust Evaluation Framework for Extreme Risk Forecasts}}

    %% AUTHORS %%%%%%%%%%%%%%%%%%%%%%%%%%%%%%%%%%%%%%%%%%%%%%%%%%%%%%%%%%%%%%%%%%%%
    %% Leave this section commented out so that the paper is blinded for review.
    %% Group authors per affiliation:
     \author{Sotirios D. Nikolopoulos}
     \address{Department of Accounting and Finance, University of Peloponnese, Greece}
%     \fntext[ssnote]{Anything you want to add as a footnote about Susan.}

%     \author {Bill Brown}
%     \address {Giggle Inc.}

    %% Only give the email address of the corresponding author
%     \cortext[cor]{Corresponding author}
     \ead{s.nikolopoulos@go.uop.gr}
    %%%%%%%%%%%%%%%%%%%%%%%%%%%%%%%%%%%%%%%%%%%%%%%%%%%%%%%%%%%%%%%%%%%%%%%%%%%%%%%%

    % ==================================================================
    % ABSTRACT
    % ==================================================================
        \begin{abstract}
            Evaluating rare-event forecasts is challenging because standard metrics collapse as event prevalence declines. Measures such as F1-score, AUPRC, MCC, and accuracy induce degenerate thresholds---converging to zero or one---and their values become dominated by class imbalance rather than tail discrimination. We develop a family of \textit{rare-event-stable} (RES) metrics whose optimal thresholds remain strictly interior as the event probability approaches zero, ensuring coherent decision rules under extreme rarity. Simulations spanning event probabilities from 0.01 down to one in a million show that RES metrics maintain stable thresholds, consistent model rankings, and near-complete prevalence invariance, whereas traditional metrics exhibit statistically significant threshold drift and structural collapse. A credit-default application confirms these results: RES metrics yield interpretable probability-of-default cutoffs (4--9\%) and remain robust under subsampling, while classical metrics fail operationally. The RES framework provides a principled, prevalence-invariant basis for evaluating extreme-risk forecasts.
        \end{abstract}
        \begin{keyword}
            Rare events\sep  Classification\sep  Decision analysis\sep  Proper scoring rules\sep  Early warning systems
            % Suggested keywords are listed at https://ijf.forecasters.org/keywords/
        \end{keyword}

    \end{frontmatter}

    % ==================================================================
    % 1. INTRODUCTION
    % ==================================================================
    \section{Introduction}

    Forecasting rare events is fundamental in credit risk, crisis prediction, fraud detection, and safety engineering. In such settings, the prevalence of positive outcomes may lie orders of magnitude below conventional levels, yet the cost of missed detections remains substantial. Although probability forecasting methods have advanced rapidly, the evaluation of these forecasts remains challenging: widely used performance metrics behave inconsistently as prevalence declines, often producing unstable or operationally difficult-to-justify decision thresholds.

    A central observation motivating this paper is that traditional threshold-based metrics such as the F1-score, MCC, and Balanced Accuracy embed marginal trade-offs that depend implicitly on the event prevalence~$\pi$. As $\pi \to 0$, these implicit trade-offs diverge, forcing the induced optimal thresholds toward extreme values even when the underlying predictive model remains unchanged. This threshold-collapse mechanism has been noted empirically but has lacked a general theoretical explanation. As rarity increases, these metrics increasingly obscure meaningful differences in tail discrimination precisely where reliable operational decision-making depends on stable threshold behaviour.

    We address this gap by developing a class of Rare-Event-Stable (RES) performance metrics that remain well behaved under extreme rarity. These metrics introduce a policy parameter~$\alpha$ that reflects the institution’s relative tolerance for false positives and false negatives. Crucially, $\alpha$ is a stable preference parameter, while the implementation threshold~$\delta^{\ast}$ remains a data-driven quantity. This separation enables coherent and interpretable decision-making across models, samples, and prevalence regimes.

    \paragraph{Contributions}
    This paper makes four primary contributions to the evaluation of rare-event classification. First, we provide an analytical explanation for the structural instabilities exhibited by common metrics such as the F1-score, MCC, and Balanced Accuracy, and we document these instabilities empirically across a wide range of prevalence conditions. Second, we develop a general class of RES metrics based on a prevalence-invariant policy parameter~$\alpha$ that separates stable institutional preferences from the data-driven determination of~$\delta^{\ast}$. Third, we propose practical calibration procedures that map historical operating policies, capacity constraints, and explicit loss structures into a transparent and interpretable choice of~$\alpha$. Fourth, through large-scale simulations and an empirical application to credit-default forecasting, we show that RES metrics yield interior, stable, and economically interpretable thresholds even under extreme rarity, in contrast to the degeneracy observed in traditional performance measures.

    The remainder of the paper is organised as follows. Section~\ref{sec:literature} reviews the literature and conceptual framework. Section~\ref{sec:foundations} establishes the decision-theoretic foundations, motivating the RES metrics introduced in Section~\ref{sec:RES}. The asymptotic behaviour is derived in Section~\ref{sec:asymptotic}, followed by simulation and empirical results in Sections~\ref{sec:simulation} and~\ref{sec:application}. Technical proofs, detailed experimental protocols, calibration procedures, and extensive supplementary diagnostics are provided in Appendices~A--J.

    % ==================================================================
    % 2. LITERATURE REVIEW AND CONCEPTUAL FRAMEWORK
    % ==================================================================

    \section{Literature Review and Conceptual Framework}
    \label{sec:literature}
    Forecast evaluation for rare events lies at the intersection of several established strands of research: the theory of proper scoring rules and probabilistic evaluation, the extensive literature on class imbalance and rare-event modelling, and the operational practice of threshold-based alarm systems used in supervisory and risk-management environments. Although these literatures are often connected implicitly in applied work, their combined implications for decision-making under extreme rarity have received comparatively little theoretical scrutiny. This section synthesizes these strands, identifies the conceptual gap that motivates our contribution, and develops the framework underlying the RES approach.

    The formal evaluation of probability forecasts typically relies on proper scoring rules such as the Brier score and the logarithmic score \citep{gneiting2007strictly, gneiting2011comparing}. These criteria are theoretically attractive because they reward full-distribution accuracy and align with Bayesian decision theory under well-specified loss functions. However, by integrating performance over the entire distribution, proper scoring rules may underweight tail behaviour, a concern raised repeatedly in risk-management and macroeconomic forecasting contexts. Studies such as \citet{diks2011likelihood} emphasise the importance of examining tail behaviour explicitly, while \citet{lahiri2013evaluating} show that standard evaluation techniques can fail to capture economic value under asymmetric loss. Experimental evidence further indicates that forecasters implicitly recognise increased tail uncertainty by widening prediction intervals in the extremes \citep{sonsino2025decrease}. These observations highlight the distinction between full-distribution evaluation and the threshold-based decisions that govern operational intervention.

    Alongside this literature, a broad methodological effort has sought to address the modelling
    challenges associated with rare events and severe class imbalance. Research in credit risk
    \citep{antunes2018forecasting, brownlees2015srisk, dinnocenzo2024modeling}, fraud detection \citep{hernandez2024financial, ngai2011application}, anomaly detection \citep{breunig2000lof, wang2020deep}, and early-warning systems \citep{candelon2014currency} has produced numerous strategies to mitigate imbalance, including resampling techniques such as SMOTE \citep{chawla2002smote, he2009learning}, rare-event correction to logistic regression \citep{king2001logistic}, and imbalance-aware learning algorithms \citep{lin2017focal}. Yet despite significant methodological progress, evaluation practice often relies on metrics whose structural behaviour under extreme imbalance remains poorly understood. The general limitations of F-measures and correlation-based metrics have been widely discussed
    \citep[e.g.,][]{powers2011evaluation, flach2015precision, saito2015precision, davis2006relationship},
    and recent analytical work \citep{minus2025behavior} demonstrates that such metrics may behave
    unpredictably as the event probability becomes very small. These findings reinforce the need for
    evaluation criteria whose optimisation behaviour remains stable and interpretable in the rare-event limit.

    The operational literature on threshold selection and alarm systems provides additional insight. Threshold rules in credit supervision, systemic-risk monitoring, fraud screening, and safety engineering commonly rely on ROC analysis, Youden’s index \citep{youden1950index}, and cost-sensitive classification frameworks \citep{hand2009measuring, hand2010proper, drummond2006roc}. This literature recognises that threshold choice must reflect an institution’s tolerance for false positives versus false negatives, and that such tolerances are often asymmetric in high-stakes settings. Echoing this view, \citet{tetlock2023false}, drawing on \citeauthor{taleb2022single} (\citeyear{taleb2022single}), argues that meaningful alarm policies require marginal trade-offs that reflect policymaker risk preferences rather than numerical artefacts of imbalance.

    Two empirical regularities emerge across these literatures. First, traditional classification metrics exhibit systematic instability as prevalence declines. Their implicit marginal penalties for false positives and false negatives vary systematically with prevalence, even when the underlying decision costs are intended to be prevalence-invariant. Empirical studies have consistently documented threshold drift, volatility, and degradation of discriminative resolution \citep{flach2015precision, saito2015precision, davis2006relationship}. The recent M6 forecasting competition \citep{makridakis2025m6} reinforces this point: even highly accurate probability forecasts can lead to poor operational decisions when the evaluation metric is misaligned with the decision environment. Second, despite extensive research on class imbalance, no widely used metric provides a structural guarantee of threshold stability as $\pi \to 0$; most either collapse to boundary thresholds or lose discriminative power, raising concerns about their suitability for operational use.

    Taken together, these findings reveal a conceptual gap: while much is known about modelling rare events and about evaluating full probability forecasts, relatively little is known about the structural behaviour of threshold-based metrics under extreme rarity. The central premise motivating our contribution is that threshold-based metrics must remain well behaved as prevalence declines; otherwise, they cannot support coherent or defensible decision policies. The RES framework introduced in the next section addresses this gap by formalizing a class of metrics whose marginal trade-offs remain stable as $\pi \to 0$, ensuring interior thresholds and consistent rankings across prevalence regimes.

    The following section formalizes these requirements and develops the structural conditions for rare-event stability.

    \section{Decision-Theoretic Foundations}
    \label{sec:foundations}

    Forecast evaluation in rare-event environments ultimately reduces to a decision problem: whether a probabilistic forecast should trigger an alarm. Institutions observing a probability estimate $\eta(x)$ must specify a threshold $\delta \in [0,1]$ and intervene whenever $\eta(x) \ge \delta$. The practical value of any evaluation metric therefore depends not only on its numerical score but on the behaviour of the metric-induced optimal threshold,
    \[
    \delta^{\ast} = \arg\max_{\delta \in [0,1]} \E[M(\delta; Y, \eta(X))],
    \]
    whose properties determine how probabilistic forecasts are translated into operational actions. This section develops the decision-theoretic foundations for analysing $\delta^{\ast}$ in the asymptotic regime where the event probability $\pi = \Pr(Y=1)$ becomes vanishingly small. We show that many widely used metrics become structurally incompatible with coherent alarm systems as prevalence declines, and we use this insight to motivate the rare-event-stable (RES) framework developed in the next section.

    Alarm thresholds play a central role in credit regulation, early-warning systems, fraud detection, and systemic-risk surveillance. For a performance metric to guide decisions in such environments, its induced threshold must satisfy two essential coherence requirements. First, the threshold must remain well defined and unique across similar environments; metrics whose maximising thresholds oscillate between extreme values produce inconsistent and operationally meaningless alarm behaviour. Second, the threshold must evolve predictably as prevalence changes. Institutions routinely operate under shifting baseline risks, and an evaluation criterion that produces erratic or discontinuous threshold movements as $\pi$ declines cannot support stable policy or risk-management decisions.

    These considerations become particularly salient in the rare-event limit, where we consider asymptotic sequences in which $\pi \to 0$ while the conditional forecast distribution remains fixed. This framework captures practical environments where events occur with probabilities far below one percent, including credit defaults in high-quality portfolios, severe system failures, and geopolitical tail risks. Two conditions are necessary for meaningful evaluation in this limit.

    \paragraph{Condition C1: Bounded Optimal Thresholds}
    A metric must produce an optimal threshold $\delta^{\ast}$ that remains strictly interior as $\pi \to 0$. Thresholds that collapse to $0$ or $1$ are operationally uninformative, since they correspond to always-alarm or never-alarm rules regardless of the model’s discriminatory content. Boundedness ensures that the metric continues to operate in the region of the score distribution where meaningful discrimination is possible.

    \paragraph{Condition C2: Stable Model Rankings}
    A metric must preserve the ordering of models as prevalence declines. If one model provides stronger upper-tail discrimination than another, the evaluation criterion must continue to reflect this as events become rarer, rather than allowing prevalence effects to dominate the metric’s behaviour.

    The structure of traditional metrics explains why these coherence requirements often fail. Metrics built from confusion-matrix components depend on $\TPR(\delta)$ and $\FPR(\delta)$, and their expected value can be expressed as a function of these quantities and of $\pi$. When $\pi$ is small, terms weighted by $(1 - \pi)$ dominate, meaning that the contribution of the negative class overwhelms that of the positive class. This imbalance causes many metrics to overweight the penalty for false positives: as prevalence declines, the marginal trade-off embedded in the metric shifts toward avoiding false alarms, forcing the optimal threshold toward the upper boundary regardless of the underlying signal strength. This mechanism is not a sampling artifact but a structural feature of the metric’s functional form.

    This observation explains the systematic collapse of widely used metric families. Precision–recall measures such as the F1-score deteriorate because precision scales with $\pi$ in a way that becomes negligible as $\pi \to 0$, producing conflicting optimisation pressures that push the threshold toward either $0$ or $1$ depending on the distributional tails. Correlation-based metrics such as the Matthews Correlation Coefficient incorporate multiplicative $\sqrt{\pi(1-\pi)}$ terms that force the maximising threshold toward $1$, effectively implementing a never-alarm rule even when models exhibit strong discriminatory power. The Area Under the ROC Curve, though threshold-free, becomes uninformative in the rare-event limit because ROC geometry is insensitive to prevalence; AUC saturates near unity and loses the ability to discriminate between models with materially different tail behaviour. Conventional accuracy measures behave similarly: as $\pi \to 0$, accuracy converges to the true negative rate, systematically ignoring the ability to detect rare events unless misclassification costs are explicitly rescaled with prevalence.

    These pathologies highlight the necessity of threshold boundedness in extreme-risk environments. A metric is rare-event-stable if its induced threshold remains strictly interior as $\pi \to 0$, thereby continuing to operate in the region of the score distribution where meaningful discrimination between rare events and non-events occurs. Metrics failing this condition cannot support consistent operational interpretation, as their induced decision rules collapse when prevalence becomes small.

    One might argue that threshold collapse is optimal under standard Bayesian decision theory with fixed misclassification costs, since the Bayes-optimal threshold depends directly on $\pi$. However, many high-stakes environments differ markedly from this classical setting. In systemic-risk monitoring, missing a single important event carries consequences far exceeding those implied by constant misclassification costs; in credit regulation, defaults in high-quality portfolios trigger supervisory attention independently of their base rate; and in reliability engineering, rare failures often entail non-linear losses. In such environments, the effective false-negative cost increases as the event becomes rarer, scaling approximately as $1/\pi$. Under this structure, the decision-theoretic optimum coincides with the rare-event-stable regime. The RES framework formalises this institutional reality by embedding prevalence-invariant marginal trade-offs through a policy parameter $\alpha$ that captures tolerance for false positives relative to false negatives, while ensuring that the induced implementation threshold remains interior as events become scarce.

    The decision-theoretic analysis therefore yields three central insights. Many classical metrics violate both boundedness and ranking stability, producing thresholds and model rankings that are driven by prevalence rather than discrimination. Rare-event stability is a structural property dictated by a metric’s functional form, not a statistical issue remediable by larger samples or alternative estimators. And in practical environments where rare events carry disproportionate consequences, evaluation metrics must embed cost structures that do not collapse with prevalence. These insights motivate the development of the RES metric family in Section~\ref{sec:RES}, where we construct evaluation criteria designed explicitly to retain their interpretability and operational meaning in the rare-event limit.

    \section{Rare-Event-Stable Metrics}
    \label{sec:RES}

    Building on the structural failures identified in Section~\ref{sec:foundations}, this section develops a class of performance metrics whose induced decision rules remain coherent as $\pi \to 0$. A metric is operationally viable in extreme-imbalance settings only if it satisfies the two rare-event requirements of bounded optimal thresholds (Condition~C1) and stable model rankings (Condition~C2). Metrics meeting these requirements, which we term \emph{rare-event-stable} (RES), maintain interpretability and preserve discrimination even when events are extraordinarily rare. In what follows, we characterise the structural form of such metrics, derive the conditions under which rare-event stability holds, introduce a canonical example, and clarify how these metrics relate to classical measures such as Balanced Accuracy and Youden’s $J$.

    Traditional evaluation metrics collapse asymptotically because their marginal penalties for false positives and false negatives diverge as $\pi$ decreases. When prevalence is small, the negative class dominates expressions involving $(1-\pi)$, leading many metrics to overweight the penalty for false positives and forcing the optimal threshold $\delta^{\ast}$ toward a boundary irrespective of the underlying signal strength. This mechanism explains why measures such as F1, MCC, and accuracy routinely generate degenerate decision rules in rare-event environments. To avoid this outcome, RES metrics must embed a trade-off between the true positive rate (TPR) and the false positive rate (FPR) that does not depend on prevalence. The implied marginal rate comparison must converge to a finite, non-zero limit as $\pi \to 0$, ensuring that $\delta^{\ast}$ remains strictly interior. Intuitively, a rare-event-stable metric must reward improvements in extreme-tail detection while penalising false positives in a manner that neither explodes nor collapses with prevalence; metrics whose trade-offs become dominated by the negative class necessarily violate threshold boundedness.

    To illustrate this structural requirement, consider metrics that can be expressed or locally approximated in the additive-discriminatory form
    \[
    \E[M(\delta)]
    =
    A(\pi)\,\TPR(\delta) - B(\pi)\,\FPR(\delta) + C(\pi),
    \qquad A(\pi)>0, \; B(\pi)>0.
    \]
    Differentiating with respect to $\delta$ yields the first-order condition
    \[
    A(\pi)\, f_1(\delta^{\ast}) = B(\pi)\, f_0(\delta^{\ast}),
    \]
    implying that the optimal threshold satisfies
    \[
    \frac{f_1(\delta^{\ast})}{f_0(\delta^{\ast})}
    =
    \frac{B(\pi)}{A(\pi)}.
    \]
    When the ratio $B(\pi)/A(\pi)$ converges to a finite, positive constant as $\pi \to 0$, the threshold $\delta^{\ast}(\pi)$ converges to an interior limit under the monotone-likelihood-ratio (MLR) assumption, thereby satisfying Condition~C1. Conversely, if this ratio diverges or vanishes, the induced threshold collapses to a boundary, rendering the metric operationally meaningless in extreme-imbalance environments.

    Metrics that satisfy rare-event stability obey three structural principles. Balanced asymptotics require that the implied trade-off between TPR and FPR remains finite as prevalence declines; no term may diverge like $1/\pi$ or vanish like $\pi$ in a manner that distorts the optimisation problem. Tail sensitivity requires that the metric increase when TPR improves at high thresholds, preserving responsiveness to the region of the distribution where rare-event discrimination occurs. Prevalence invariance requires that the effective penalty on false positives not scale implicitly with the size of the negative class; otherwise, the induced threshold necessarily collapses to a boundary. Metrics obeying these principles avoid the degeneracies highlighted in Section~\ref{sec:foundations} and therefore satisfy Conditions~C1 and C2. This characterisation offers a practical diagnostic tool, since many classical metrics violate at least one of these principles, explaining their empirical instability in rare-event applications.

    A transparent and analytically tractable representative of the RES family is
    \begin{equation}
        M_{\mathrm{RE}}(\delta)
        =
        \frac{\TPR(\delta)}
        {\alpha\,\FPR(\delta) + (1 - \alpha)},
        \qquad
        \alpha \in (0, 1).
        \label{eq:MRE}
    \end{equation}
    This metric possesses three essential properties. First, it is prevalence independent because the denominator contains no prevalence term, ensuring that the marginal trade-off between TPR and FPR remains stable as $\pi$ decreases. Second, it induces an interior optimal threshold: the maximiser satisfies a likelihood-ratio condition with a finite right-hand side, guaranteeing convergence to a non-degenerate limit and fulfilling Condition~C1. Third, its optimisation depends only on the pair $(\TPR,\FPR)$, ensuring that model rankings remain consistent across prevalence regimes and satisfying Condition~C2. The parameter $\alpha$ serves as an interpretable policy lever, encoding the institution’s tolerance for false positives relative to false negatives while preserving rare-event stability across the entire preference spectrum. Further discussion of the policy parameter $\alpha$, including practical calibration procedures and operational interpretations, is provided in Appendix~D.

    \paragraph{Generalizing the RES Family}
    While $M_{\mathrm{RE}}$ defined in \eqref{eq:MRE} serves as the canonical and most tractable representative, the RES framework encompasses a broader family of functional forms. Any metric $M(\delta)$ can be classified as Rare-Event-Stable provided its marginal rate of substitution between TPR and FPR converges to a finite non-zero constant as $\pi \to 0$. For example, a generalized family could take the form:
    \begin{equation}
        M_{\mathrm{gen}}(\delta) = \frac{\TPR(\delta)^{\gamma}}{\alpha \FPR(\delta) + (1-\alpha)}, \quad \gamma > 0.
    \end{equation}
    Here, $\gamma$ modulates the curvature of the indifference curves in ROC space. We focus on the linear case ($\gamma=1$) in this paper because of its direct link to cost-sensitive Bayesian decision theory, but non-linear extensions may offer additional flexibility for institutions with varying risk aversion in the extreme tail.

    The structure of $M_{\mathrm{RE}}$ connects naturally to cost-sensitive accuracy. Classical Bayesian decision theory \citep{elkan2001foundations} minimises the loss
    \[
    L = \pi\, C_{FN}(1-\TPR) + (1-\pi)\, C_{FP}\,\FPR.
    \]
    With fixed misclassification costs, the term $(1-\pi) C_{FP}\FPR$ dominates as $\pi \to 0$, forcing the optimal threshold toward $1$ unless costs are explicitly rescaled with prevalence. In many high-stakes settings, however, the cost of missing a rare event increases sharply as the event becomes rarer, effectively scaling as $C_{FN} \propto 1/\pi$. Under such conditions, the marginal trade-off embedded in $M_{\mathrm{RE}}$ corresponds precisely to the implicit institutional loss structure, providing a decision-theoretic foundation for the RES formulation and clarifying why prevalence-invariant marginal penalties are essential for coherent alarm rules in extreme-risk environments.

    Classical metrics such as Balanced Accuracy and Youden’s $J$ satisfy Condition~C1 because their optimal thresholds solve the likelihood-ratio equation $f_1(\delta^{\ast}) = f_0(\delta^{\ast})$, corresponding to an implicit cost ratio of $1{:}1$. They are therefore technically rare-event-stable. However, they lack tunability, since many institutions require asymmetric weighting of false negatives and false positives that these metrics cannot express. RES metrics such as $M_{\mathrm{RE}}$ generalise these classical measures by introducing a policy parameter while preserving threshold boundedness.

    Rare-event-stable metrics offer several advantages in extreme-risk environments. Their induced thresholds reflect stable institutional preferences rather than numerical artefacts of prevalence, ensuring operational alignment across models and samples. Because their marginal trade-offs remain invariant to the size of the negative class, they preserve the meaning of evaluation even under severe imbalance. Their tail sensitivity guarantees that improvements in upper-tail discrimination translate into substantive gains in the evaluation criterion, rather than being dominated by fluctuations in the non-event distribution. Finally, they maintain stable model comparison by avoiding the distortions inherent in metrics whose values are driven primarily by prevalence. These properties make RES metrics suitable for high-stakes applications such as credit-default forecasting, anomaly detection, industrial fault prediction, and systemic-risk surveillance, where traditional metrics routinely fail as events become extremely rare.

    Practical procedures for calibrating $\alpha$ under different institutional settings are presented in  Appendix~E (Calibration of the Policy Parameter $\alpha$).

    % ==================================================================
    % 5. ASYMPTOTIC BEHAVIOR AND ROBUSTNESS CONDITIONS
    % ==================================================================

    \section{Asymptotic Behavior and Robustness Conditions}
    \label{sec:asymptotic}

    This section formalizes the asymptotic behaviour of rare-event-stable (RES) metrics as the event probability $\pi$ approaches zero. Our aim is to characterise the limiting form of the metric-induced optimal threshold, establish conditions under which this threshold remains interior, and contrast these properties with the systematic degeneracies exhibited by widely used performance measures. The analysis connects directly to the coherence requirements introduced in Section~\ref{sec:foundations}, namely bounded optimal thresholds (Condition C1) and stable model rankings (Condition C2).

    \subsection{Preliminaries and Conditional Distribution Structure}

    Let the conditional forecast distributions be given by
    \[
    F_1(t) = \Pr(\eta(X) \le t \mid Y=1),
    \qquad
    F_0(t) = \Pr(\eta(X) \le t \mid Y=0),
    \]
    with associated densities $f_1(t)$ and $f_0(t)$ defined on $(0,1)$. The likelihood ratio
    \[
    \Lambda(t) = \frac{f_1(t)}{f_0(t)}
    \]
    plays a central role in determining optimal thresholds. Throughout this section we impose a monotone-likelihood-ratio (MLR) condition, which guarantees uniqueness of the induced optimal threshold. Under mild regularity conditions, the functions $\TPR(\delta)$ and $\FPR(\delta)$ are continuously differentiable in $\delta$, permitting the use of standard first-order optimality arguments.

    \subsection{Limiting Form of the Optimal Threshold for RES Metrics}

    Consider a RES metric whose expected value can be represented as
    \[
    \E[M(\delta)]
    =
    A(\pi)\,\TPR(\delta)
    -
    B(\pi)\,\FPR(\delta)
    +
    C(\pi),
    \]
    with $A(\pi) > 0$ and $B(\pi) > 0$. Differentiating with respect to $\delta$ yields the first-order condition
    \[
    A(\pi)\, f_1(\delta^{\ast}) = B(\pi)\, f_0(\delta^{\ast}),
    \]
    so that the optimal threshold satisfies the likelihood-ratio equation
    \[
    \Lambda(\delta^{\ast}) = \frac{B(\pi)}{A(\pi)}.
    \]
    A metric is rare-event-stable if and only if the marginal penalty ratio converges to a finite, positive limit:
    \begin{equation}
        0
        <
        \lim_{\pi\to 0}\frac{B(\pi)}{A(\pi)}
        <
        \infty.
        \label{eq:ratio_condition}
    \end{equation}
    Under the MLR condition, the induced threshold therefore converges to a unique interior point $\delta_{\infty}$ that solves
    \[
    \Lambda(\delta_{\infty}) = \lim_{\pi\to 0}\frac{B(\pi)}{A(\pi)}.
    \]
    This captures the bounded-threshold requirement (Condition C1) and shows that RES metrics maintain coherent decision rules even when $\pi$ becomes arbitrarily small.

    \subsection{Robustness to Reductions in Event Probability}

    Metrics satisfying \eqref{eq:ratio_condition} possess several forms of robustness as prevalence declines. First, threshold continuity holds: if the ratio $B(\pi)/A(\pi)$ varies smoothly in $\pi$, then the induced threshold $\delta^{\ast}(\pi)$ varies smoothly as well, ensuring predictable behaviour under changing imbalance conditions, specifically prior probability shift \citep{moreno2012unifying}. Second, preservation of model ordering is guaranteed: if model A dominates model B in likelihood-ratio ordering, then $M_A(\pi)$ exceeds $M_B(\pi)$ for all sufficiently small $\pi$, thereby satisfying Condition C2. Third, RES metrics exhibit resilience in finite samples, since their dependence on the pair $(\TPR,\FPR)$ rather than on raw event counts enables stable behaviour even when the number of observed positive cases is extremely small.

    \paragraph{Relaxation of the MLR Assumption}
    The asymptotic analysis relies on the Monotone Likelihood Ratio (MLR) property to guarantee the uniqueness of the optimal threshold $\delta^{\ast}$. In practice, highly complex models such as deep neural networks may produce calibrated probabilities that do not strictly satisfy MLR, potentially resulting in a likelihood ratio $\Lambda(\delta)$ that is non-monotonic. It is important to note that the rare-event stability of RES metrics does not depend on uniqueness. Even if $\Lambda(\delta)$ oscillates, the condition for stability (Equation \ref{eq:ratio_condition}) ensures that \emph{all} local maxima of the RES metric remain strictly interior as $\pi \to 0$. In contrast, traditional metrics such as the F1-score suffer from structural collapse where the optimality condition diverges to the boundary regardless of the shape of $\Lambda(\delta)$. Therefore, while MLR violation may introduce local optima, it does not compromise the fundamental resistance of RES metrics to prevalence-driven threshold collapse.

    \subsection{Why Standard Metrics Fail Asymptotically}

    The asymptotic framework also clarifies the structural failures of widely used metrics. Precision–recall measures such as F1 and $F_{\beta}$ rely on
    \[
    \mathrm{Prec}(\delta)
    =
    \frac{\pi\,\TPR(\delta)}
    {\pi\,\TPR(\delta) + (1-\pi)\,\FPR(\delta)},
    \]
    and as $\pi \to 0$, the denominator becomes dominated by $(1-\pi)\,\FPR(\delta)$, causing precision to collapse unless $\FPR(\delta)$ shrinks at an extreme rate. Maximising F1 therefore drives $\delta^{\ast}$ toward zero to inflate recall or toward one to inflate precision, depending on the tail shape of the score distribution. This instability reflects the fact that F1 induces a prevalence-dependent implicit cost ratio, a structural incoherence formally derived by \citet{hand2018note}. Either outcome violates Conditions C1 and C2. Correlation-based metrics such as MCC contain multiplicative $\sqrt{\pi(1-\pi)}$ factors that direct optimisation toward $\delta^{\ast}\to 1$, effectively producing a never-alarm rule even for strongly discriminative models and thereby violating threshold boundedness. Accuracy and fixed-cost variants suffer an analogous failure: since $\mathrm{Acc} = \pi\,\TPR + (1-\pi)\,\TNR$, accuracy converges to $\TNR$ as $\pi\to 0$ and thus ignores upper-tail discrimination altogether, violating ranking stability. Although AUC does not involve an explicit threshold and therefore does not violate Condition C1, ROC geometry becomes dominated by the negative class as prevalence decreases. AUC saturates near unity even when models differ substantially in tail behaviour, violating Condition C2 and rendering the metric uninformative in rare-event environments. These failures are structural, arising from the divergence or vanishing of $B(\pi)/A(\pi)$, and not merely consequences of sampling variability.

    \subsection{Practical Implications}

    The asymptotic analysis yields several practical implications for evaluation under extreme rarity. RES metrics induce stable thresholds that remain coherent even when the event probability is extremely small, ensuring operational interpretability. They support consistent model comparison by maintaining meaningful performance rankings across prevalence regimes and avoiding the distortions that plague classical measures. They offer robustness under data scarcity because they depend on the shape of the forecast distribution in the extreme upper tail rather than on the absolute number of observed events. Finally, they provide tail-sensitive evaluation: improvements in the region of the distribution most relevant for detecting rare events translate directly into metric gains, aligning evaluation with institutional priorities. These asymptotic properties underpin the empirical findings in Sections~\ref{sec:simulation} and~\ref{sec:application}, where RES metrics consistently display stability and interpretability in sharp contrast to the collapse exhibited by traditional measures. Additional robustness checks and supplementary asymptotic diagnostics appear in Appendices~B and~E.

    % ==================================================================
    % 6. SIMULATION DESIGN AND RESULTS
    % ==================================================================

    \section{Simulation Design and Results}
    \label{sec:simulation}

    The simulation study evaluates the behaviour of rare-event-stable (RES) metrics under controlled conditions spanning five orders of magnitude in prevalence. We consider two signal regimes defined by the degree of distributional overlap between classes: a \emph{Moderate} regime (implied $\mathrm{AUC} \approx 0.90$), representing standard credit-scoring performance, and a \emph{Strong} regime (implied $\mathrm{AUC} > 0.99$), representing near-ideal separation often seen in distinct failure-mode detection. The data-generating processes and metric calculation procedures are detailed in Appendix~B. Here we report the main results, focusing on threshold drift, discriminative resolution, and the stability properties of RES metrics.

    Detailed descriptions of the simulation design, including sampling procedures, parameter settings, and implementation details, are provided in Appendix~C.

    Prevalence is varied exogenously. For each target value $\pi \in \{10^{-2},10^{-3},10^{-4},10^{-5},10^{-6}\}$, stratified samples are constructed by fixing the number of positive observations and choosing the number of negatives implied by the selected prevalence. At the most extreme rarity level, such as $\pi = 10^{-6}$, this procedure yields samples containing, for example, $20$ positive and $20$ million negative observations, capturing the geometry of the rare-event limit without introducing additional resampling noise. Across all regimes, $2{,}000$ independent replications are generated. For every metric, discrimination regime, and prevalence level, the optimal threshold $\delta^\ast$ is computed exactly by tracing the full confusion-matrix path. This design enables a direct examination of whether traditional metrics undergo threshold instability as $\pi \to 0$, whether RES metrics maintain stable interior thresholds, and whether discrimination-based ranking is preserved as prevalence varies.

   % ------------------------------------------------------------------
   \subsection{Instability of Classical Metrics}
   % ------------------------------------------------------------------

   Traditional measures exhibit substantial and systematic deterioration as prevalence declines. Table~\ref{tab:sim_ranges} summarises the range of optimal thresholds for the F1-score, Balanced Accuracy (BA), the Matthews Correlation Coefficient (MCC), and AUC across five orders of magnitude in~$\pi$. Even in this fully controlled experimental setting, large shifts in~$\delta^\ast$ appear, indicating that threshold instability is a structural property of these metrics rather than an artefact of sampling variation.

   \begin{table}[ht]
       \centering
       \caption{Optimal threshold ranges for traditional metrics across five prevalence levels ($10^{-2}$ to $10^{-6}$).}
       \label{tab:sim_ranges}
       \small
       \begin{tabular}{lcccc}
           \hline
           Model & F1 Range & BA Range & MCC Range & AUC Range \\
           \hline
           Moderate & 0.552 -- 0.992 & 0.242 -- 0.599 & 0.532 -- 0.992 & 0.879 -- 0.999 \\
           Strong   & 0.426 -- 0.842 & 0.223 -- 0.776 & 0.426 -- 0.842 & 0.996 -- 1.000 \\
           \hline
       \end{tabular}
   \end{table}

\begin{figure}[htbp]
       \centering
       \includegraphics[width=0.48\textwidth]{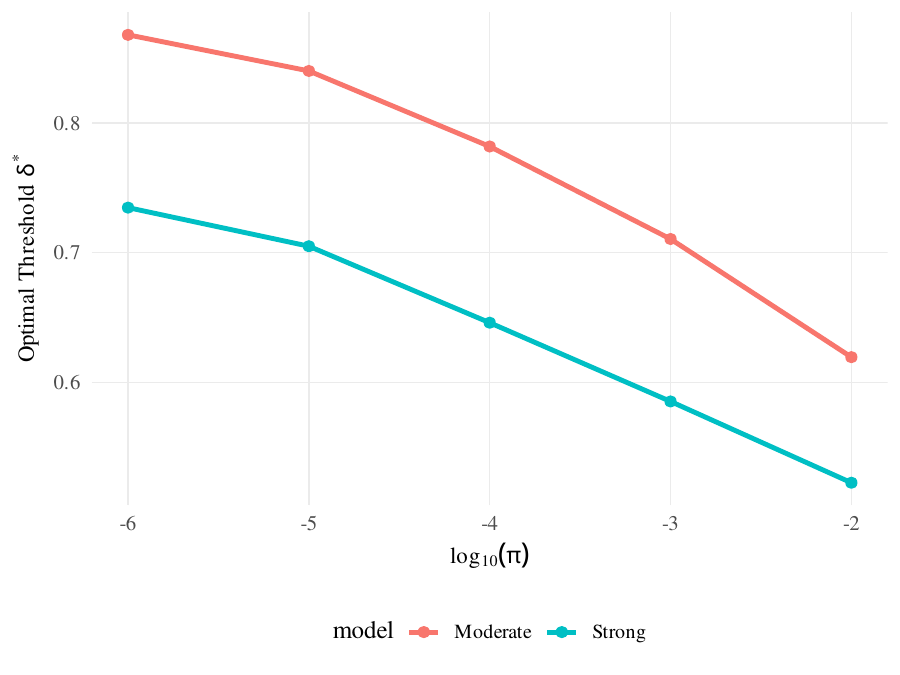}
       \includegraphics[width=0.48\textwidth]{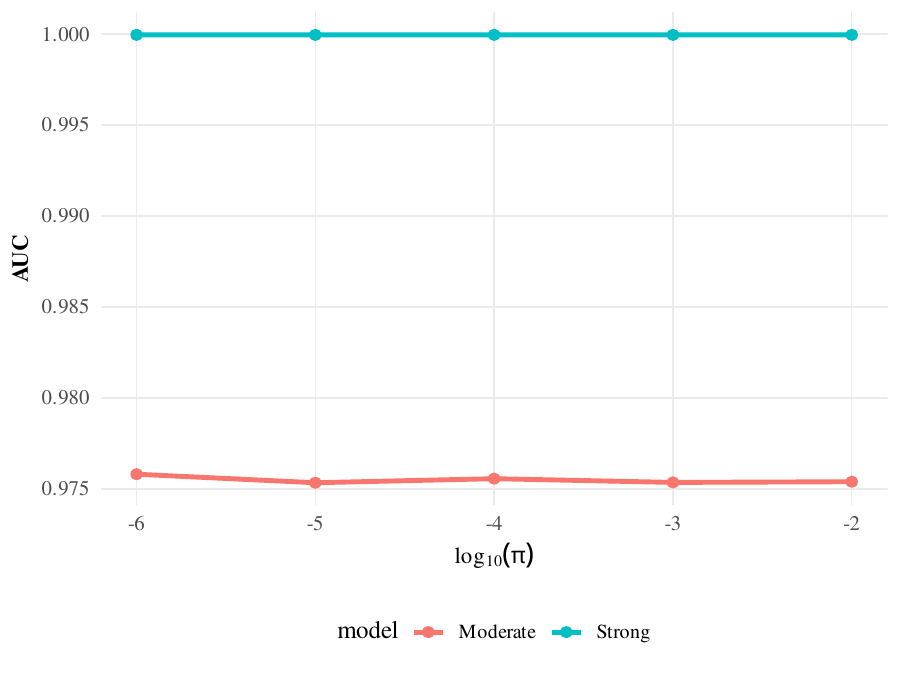}
       \caption{Behaviour of traditional metrics under extreme rarity. Left: F1-score thresholds increase monotonically as prevalence decreases ($p<0.001$). Right: AUC saturates near $1.0$, losing discriminatory power in the extreme tail.}
       \label{fig:f1_auc}
   \end{figure}

   The F1-score displays pronounced threshold drift: in the Moderate regime, the optimal threshold moves from interior values (around $0.55$) toward the upper boundary (around $0.99$) as prevalence declines. While MCC is frequently recommended as a reliable, informative alternative to F1 in imbalanced settings \citep{chicco2020advantages}, our results indicate that this robustness does not extend to the rare-event limit. MCC exhibits a nearly identical migration to F1, with its optimal threshold spanning $0.53$--$0.99$ in the Moderate regime and $0.43$--$0.84$ in the Strong regime. Balanced Accuracy also shows substantial movement, with shifts exceeding $0.55$ under Strong separation. These changes are operationally significant: a classifier calibrated for a moderately imbalanced environment would adopt nearly opposite intervention rules under extreme rarity. The monotonic nature of this migration is illustrated in Figure~\ref{fig:f1_auc}(left).

   Table~\ref{tab:sim_collapse} reports the statistical significance of the relationship between the optimal threshold $\delta^\ast$ (or the AUC value) and $\log_{10}(\pi)$. All correlations are highly significant ($p<0.001$), confirming that the observed instability arises from the underlying metric structure rather than simulation noise.

   \begin{table}[ht]
       \centering
       \caption{Significance tests for metric instability (Spearman correlation with $\log_{10}(\pi)$).}
       \label{tab:sim_collapse}
       \small
       \begin{tabular}{lcccc}
           \hline
           Model & F1 ($p$) & BA ($p$) & MCC ($p$) & AUC ($p$) \\
           \hline
           Moderate & $<0.001$ & $<0.001$ & $<0.001$ & $<0.001$ \\
           Strong   & $<0.001$ & $<0.001$ & $<0.001$ & $<0.001$ \\
           \hline
       \end{tabular}
   \end{table}

   Although AUC does not induce a decision threshold directly, its behaviour illustrates the loss of discriminative resolution in the extreme tail. In the Strong regime, the AUC range is $0.996$--$1.000$ (Table~\ref{tab:sim_ranges}). While AUC remains a robust metric for global ranking across the entire distribution, this saturation effect renders it uninformative for distinguishing between models specifically in the rare-event limit (Figure~\ref{fig:f1_auc}, right). The combined evidence confirms that standard metrics either collapse to boundary thresholds or lose discriminative power as $\pi \to 0$, matching the structural failures predicted by the analytical results.

    % ------------------------------------------------------------------
    \subsection{Stability of RES Metrics}
    % ------------------------------------------------------------------

    The RES metrics behave in a manner fully aligned with theoretical expectations. Their induced thresholds remain interior across all prevalence levels, vary smoothly as $\pi$ changes, and maintain a clear distinction between the Moderate and Strong discrimination regimes. Table~\ref{tab:sim_res} reports mean thresholds and coefficients of variation across the full range of~$\pi$.

    \begin{table}[ht]
        \centering
        \caption{Stability of RES metric thresholds across five prevalence levels.}
        \label{tab:sim_res}
        \small
        \begin{tabular}{lcccc}
            \hline
            Model & $\alpha$ & Mean $\delta^{\ast}$ & MRE CV & Threshold CV \\
            \hline
            Moderate & 0.10 & 0.289 & 0.014 & 0.213 \\
            Moderate & 0.25 & 0.335 & 0.026 & 0.151 \\
            Moderate & 0.50 & 0.391 & 0.041 & 0.108 \\
            Strong   & 0.10 & 0.471 & 0.001 & 0.193 \\
            Strong   & 0.25 & 0.472 & 0.001 & 0.189 \\
            Strong   & 0.50 & 0.476 & 0.003 & 0.178 \\
            \hline
        \end{tabular}
    \end{table}

    RES thresholds exhibit strong prevalence invariance. In the Strong regime with $\alpha = 0.50$, the mean threshold is approximately $0.476$ across six orders of magnitude in prevalence. The coefficients of variation are far below those observed for classical metrics, indicating that RES thresholds remain stable and do not drift toward boundary behaviour as $\pi$ declines. This stability reflects the prevalence-invariant marginal trade-off embedded in the RES formulation.

    The parameter $\alpha$ consistently encodes institutional risk preferences. Lower values generate conservative thresholds with fewer alarms, whereas higher values increase sensitivity. Importantly, these relationships persist across all levels of prevalence, demonstrating that RES metrics transmit policy preferences coherently even in extreme rarity.

    \begin{figure}[htbp]
        \centering
        \includegraphics[width=0.7\textwidth]{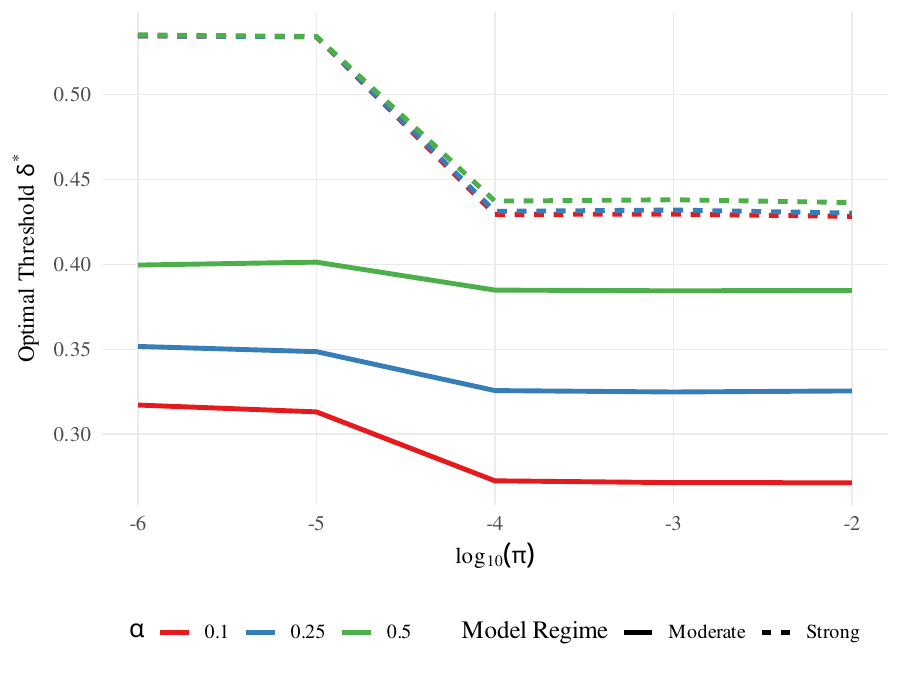}
        \caption{Stability of RES thresholds. Optimal thresholds remain interior and nearly constant across all prevalence levels. The parameter $\alpha$ modulates preferences while preserving threshold stability.}
        \label{fig:res_stability}
    \end{figure}

    Overall, the RES metrics satisfy both threshold boundedness and ranking stability. Their simulated behaviour mirrors the analytical characterisation developed in earlier sections and contrasts sharply with the structural collapse observed for traditional measures.

    % ------------------------------------------------------------------
    \subsection{Summary of Simulation Findings}
    % ------------------------------------------------------------------

    The simulation results reveal a clear and systematic distinction between the behaviour
    of traditional performance metrics and that of the proposed RES metrics under rare-event
    scaling. Classical measures such as the F1-score, Balanced Accuracy, and MCC exhibit
    substantial threshold drift as $\pi$ decreases, with optimal thresholds migrating
    towards boundary values and, in many cases, converging to near-degenerate operating
    regions. Their behaviour is monotone, statistically significant across all regimes, and
    consistent with the structural collapse predicted by the theoretical analysis.

    AUC behaves differently but suffers an analogous pathology: as prevalence declines,
    its value saturates near one, eliminating meaningful discriminative resolution in the
    extreme tail. This saturation violates ranking stability, causing models with materially
    different tail geometry to appear indistinguishable when evaluated by AUC.

    In contrast, RES metrics maintain interior, stable, and prevalence-invariant optimal
    thresholds. Across five orders of magnitude in $\pi$, RES thresholds exhibit limited
    dispersion, smooth adjustment across regimes, and negligible variability in metric
    values. The parameter $\alpha$ consistently encodes institutional preferences without
    interacting with prevalence, demonstrating that the RES framework preserves both the
    interpretation and the operational meaning of model evaluation in rare-event settings.

    Overall, the simulation evidence confirms that classical metrics undergo structural
    instability as $\pi \to 0$, whereas RES metrics satisfy both threshold boundedness and
    ranking stability. These properties establish RES metrics as reliable evaluation tools
    in extreme imbalance environments and motivate their use in the empirical application
    that follows.

    Extended tables and supplementary simulation diagnostics appear in Appendix~G; full empirical tables: Appendix F.

    % ==================================================================
    % 7. APPLICATION: CREDIT-DEFAULT FORECASTING UNDER EXTREME RARITY
    % ==================================================================

    \section{Application: Credit-Default Forecasting Under Extreme Rarity}
    \label{sec:application}

    This section evaluates the empirical performance of the RES metrics in a real forecasting environment using the ``Give Me Some Credit'' dataset from Kaggle, a widely used consumer-credit panel containing borrower characteristics and a binary indicator of serious delinquency within two years. The empirical design follows standard practice in the forecast-evaluation literature. The dataset and forecasting model are described, controlled prevalence regimes are constructed to emulate increasing rarity, and the behaviour of threshold-based metrics is analysed across these regimes. The results demonstrate that the structural instabilities identified in the theoretical and simulation analyses
    manifest clearly in real data, whereas the RES framework maintains stability and interpretability throughout.

    \subsection{Data and Model Specification}

    The dataset includes borrower-level financial and demographic variables alongside a binary default indicator. After standard preprocessing and median imputation, a LightGBM model is trained on a stratified 70\% subsample, with the remaining 30\% held out for evaluation. The model produces calibrated probability-of-default forecasts $\widehat{p}_i$, which remain fixed across all experiments so that variation in metric behaviour reflects solely the evaluation criteria rather than changes in model specification. Further descriptive statistics and preprocessing details appear in Appendix~D (Application: Data \& Model).

    The natural prevalence of default in the test sample is 6.64\%. To assess threshold behaviour under increasing rarity, additional evaluation regimes are created by down-sampling positive observations while retaining all negatives. This design preserves the geometry of low-prevalence environments without introducing artificial resampling artefacts, mirroring the `Low Default Portfolio' (LDP) problem common in regulatory risk modelling \citep{pluto2011estimating}. Table~\ref{tab:B1} summarises the resulting sample sizes.

    \begin{table}[ht]
        \centering
        \caption{Construction of prevalence regimes used in the empirical analysis.}
        \label{tab:B1}
        \begin{tabular}{lrrrr}
            \hline
            Target $\pi$ & Empirical $\pi$ & $N$ & Positives & Negatives \\
            \hline
            0.0010 & 0.0010 & 42056 & 42 & 42014 \\
            0.0050 & 0.0050 & 42225 & 211 & 42014 \\
            0.0100 & 0.0100 & 42438 & 424 & 42014 \\
            0.0200 & 0.0200 & 42871 & 857 & 42014 \\
            0.0664 & 0.0664 & 45000 & 2986 & 42014 \\
            \hline
        \end{tabular}
    \end{table}

    \subsection{Bootstrap Design}

    Within each prevalence regime, $500$ bootstrap samples of fixed size are drawn
    (\citealp{efron1993bootstrap}). For each replication, the predicted probabilities are sorted in descending order to trace the full confusion-matrix path as a function of the threshold $\delta$. This procedure yields the exact threshold $\delta^{\ast}$ that maximises each metric in each bootstrap replication. Repeating this process across samples produces empirical distributions of optimal thresholds and metric values, enabling a clean distinction between structural behaviour and sampling variability.
    Summary statistics for these distributions appear in Tables~\ref{tab:C1}--\ref{tab:C2}.

    \subsection{Threshold Behaviour of Classical Metrics}

\begin{figure}[ht]
        \centering
        \includegraphics[width=0.85\linewidth]{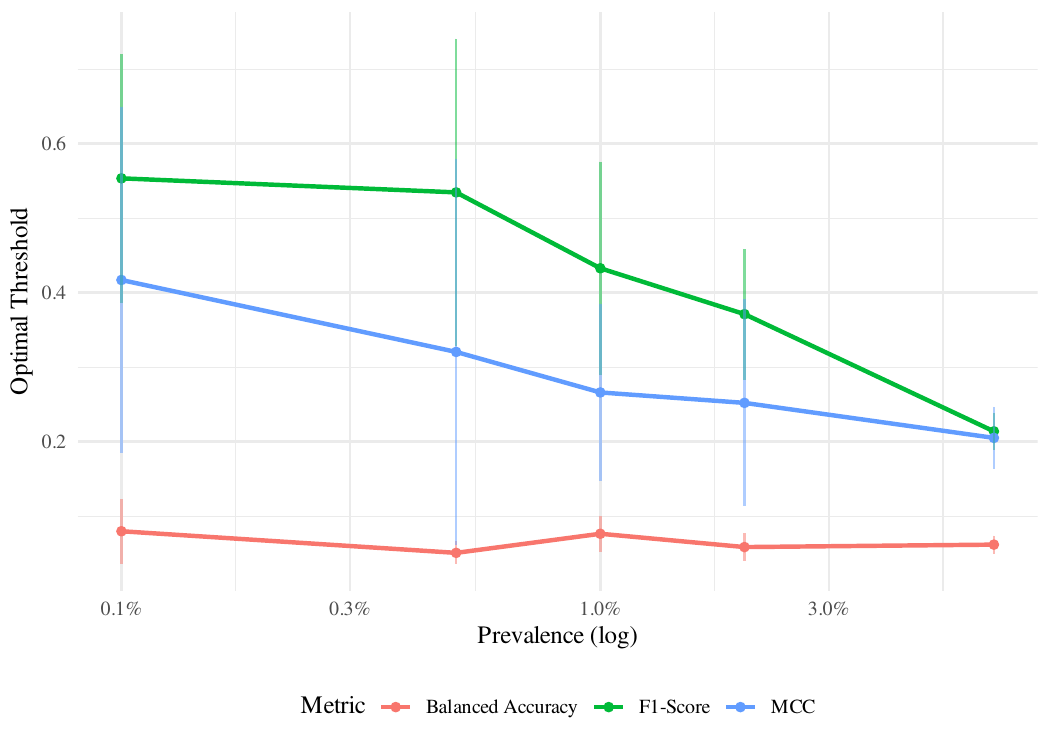}
        \caption{Optimal thresholds $\delta^{\ast}$ for F1, MCC, and Balanced Accuracy across
            prevalence regimes. Thresholds drift sharply upward as prevalence declines,
            revealing structural instability.}
        \label{fig:D1}
    \end{figure}

    Traditional metrics exhibit substantial and systematic threshold drift as prevalence declines. Figure~\ref{fig:D1} shows that the optimal thresholds for F1, MCC, and Balanced Accuracy increase monotonically with rarity, with especially large movements in the rarest regime ($\pi=0.001$). This pattern is fully consistent with the threshold-collapse mechanism predicted by the theoretical framework.

    Table~\ref{tab:C1} quantifies this instability. F1 and MCC exhibit threshold ranges exceeding 0.55 and 0.64 respectively, with coefficients of variation between 0.34 and 0.40. These imply large and economically significant shifts in the alarm policy across regimes. Balanced Accuracy is numerically stable but effectively rigid, consistently selecting a threshold between 2\% and 19\%, which limits its relevance when false positives and false negatives carry asymmetric costs.

    \begin{table}[ht]
        \centering
        \caption{Threshold stability of traditional metrics across prevalence regimes.}
        \label{tab:C1}
        \begin{tabular}{lrrrr}
            \hline
            Metric & Min $\delta^\ast$ & Max $\delta^\ast$ & Range & CV \\
            \hline
            Balanced Accuracy & 0.02 & 0.19 & 0.17 & 0.26 \\
            F1--Score         & 0.16 & 0.71 & 0.55 & 0.34 \\
            MCC               & 0.07 & 0.71 & 0.64 & 0.40 \\
            \hline
        \end{tabular}
    \end{table}

\begin{figure}[ht]
        \centering
        \includegraphics[width=0.85\linewidth]{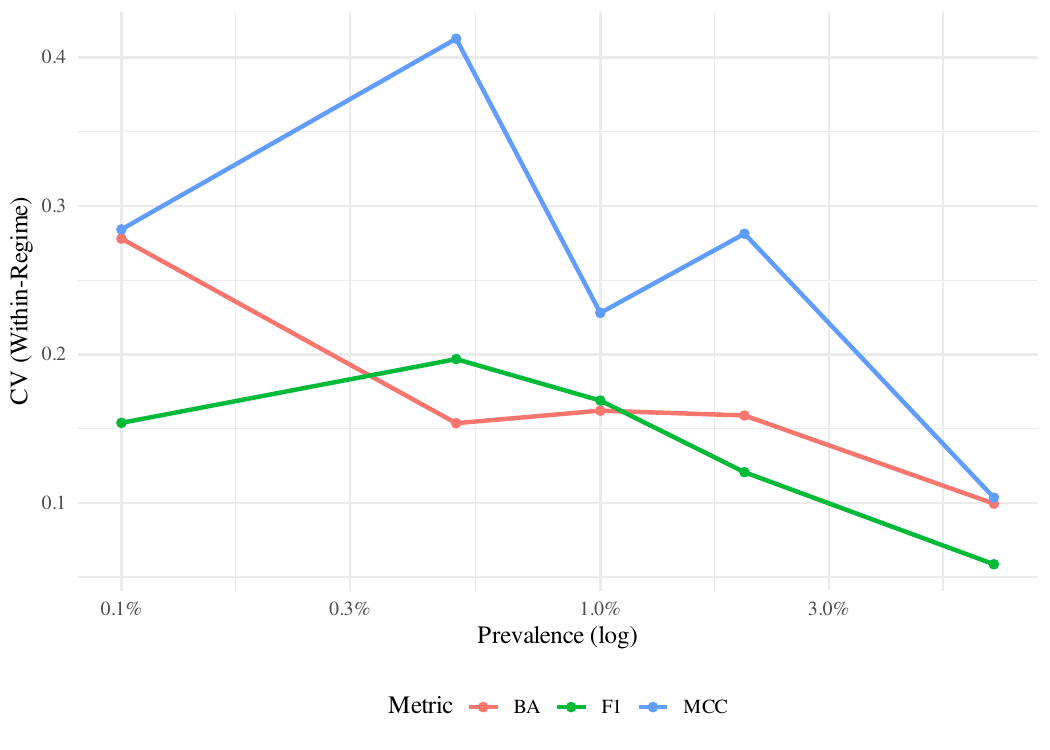}
        \caption{Coefficient of variation of optimal thresholds within each prevalence
            regime. Classical metrics exhibit significant within-regime volatility.}
        \label{fig:D3}
    \end{figure}

    Instability is also evident within each prevalence level. Figure~\ref{fig:D3} shows that both F1 and MCC exhibit substantial within-regime dispersion, indicating limited guidance regarding a stable implementation threshold even when the underlying forecast distribution is unchanged.

\begin{figure}[ht]
        \centering
        \includegraphics[width=0.75\linewidth]{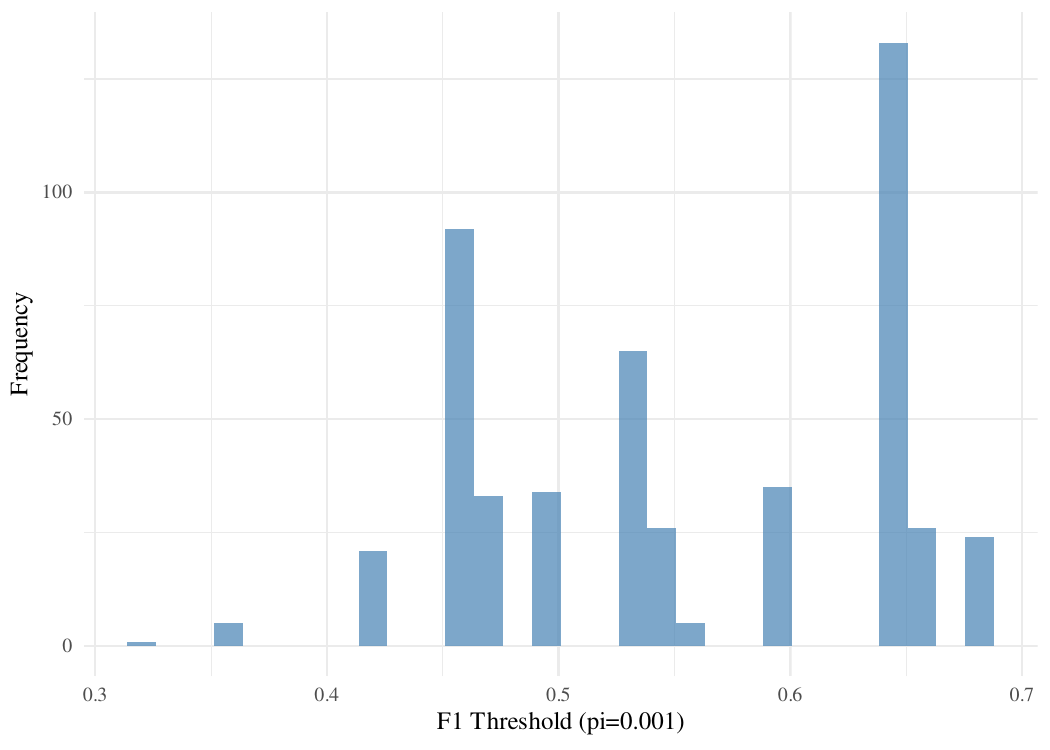}
        \caption{Distribution of F1-optimal thresholds at $\pi \approx 0.001$. The wide dispersion illustrates pronounced instability in extremely rare-event settings.}
        \label{fig:F1hist}
    \end{figure}

    At $\pi \approx 0.001$, the volatility becomes extreme. Figure~\ref{fig:F1hist} shows the bootstrap distribution of F1-optimal thresholds, revealing wide dispersion and many replications exceeding 0.70—an operationally implausible pattern for credit-risk screening.

    \subsection{Stability of RES Metrics}

\begin{figure}[ht]
        \centering
        \includegraphics[width=0.85\linewidth]{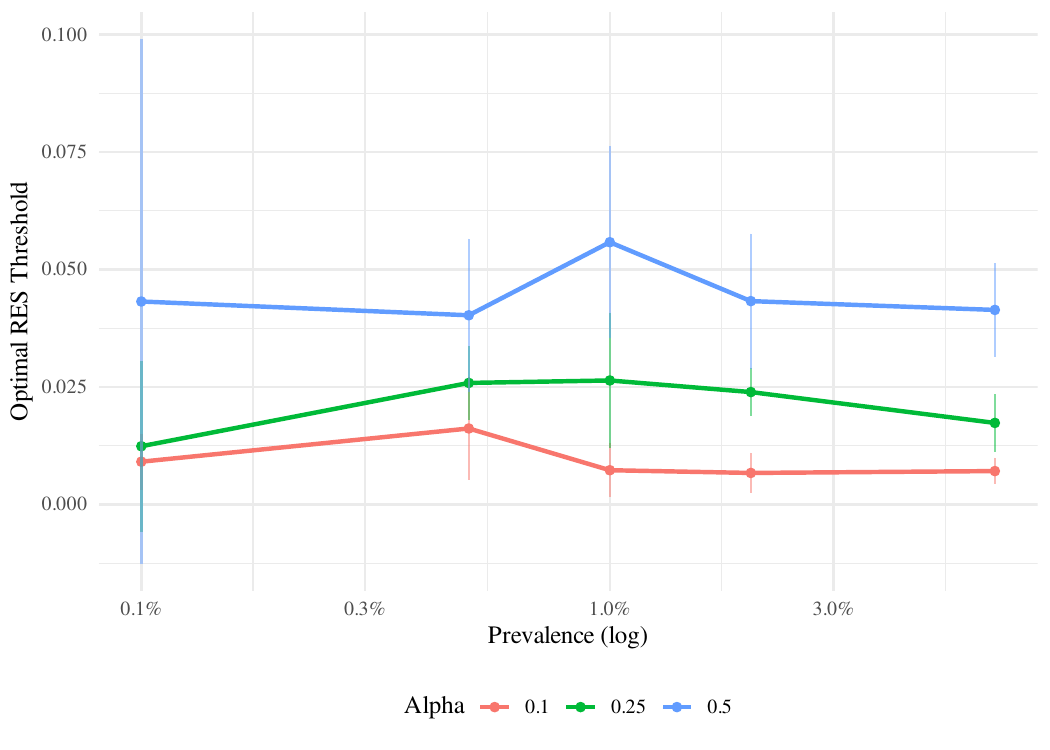}
        \caption{Optimal RES thresholds $\delta^{\ast}$ for $\alpha\in\{0.10,0.25,0.50\}$.
            Thresholds remain interior and stable even under extreme rarity.}
        \label{fig:D2}
    \end{figure}

    The RES metrics exhibit fundamentally different behaviour. As shown in
    Figure~\ref{fig:D2}, their optimal thresholds remain interior and stable across all prevalence levels. Thresholds adjust smoothly with $\alpha$ and preserve their relative ordering across regimes, reflecting consistent transmission of institutional preferences.

    Table~\ref{tab:C2} summarises these results. Across the full range of $\alpha \in \{0.10, 0.25, 0.50\}$, mean RES-threshold values remain strictly interior, lying between 1\% and 4\%. Although the coefficient of variation is higher for $\alpha = 0.10$ (CV = 0.52) due to increased sensitivity in the extreme tail, the absolute location of the thresholds still defines economically interpretable operating regions across all prevalence levels. This behaviour stands in sharp contrast to the boundary collapse observed for classical metrics.

    \begin{table}[ht]
        \centering
        \caption{RES metric performance across prevalence regimes.}
        \label{tab:C2}
        \begin{tabular}{rrrr}
            \hline
            $\alpha$ & Mean $\delta^\ast$ & Threshold CV & MRE CV \\
            \hline
            0.10 & 0.01 & 0.52 & 0.01 \\
            0.25 & 0.02 & 0.38 & 0.02 \\
            0.50 & 0.04 & 0.35 & 0.05 \\
            \hline
        \end{tabular}
    \end{table}

    \subsection{Operational Implications}

    Table~\ref{tab:E1} aggregates the empirical stability results into operational
    categories. F1 and MCC require large, prevalence-driven adjustments in $\delta^{\ast}$, leading to unstable and difficult-to-defend alarm rules. Balanced Accuracy is stable but rigid and suitable only for symmetric-cost environments. RES metrics exhibit both stability and flexibility, allowing $\alpha$ to encode institutional preferences while preserving stable operating points.

    \begin{table}[ht]
        \centering
        \caption{Operational implications of threshold behaviour across metrics.}
        \label{tab:E1}
        \begin{tabular}{lrrl}
            \hline
            Metric & Range & CV & Recommendation \\
            \hline
            Balanced Accuracy & 0.174 & 0.26 & Symmetric Only \\
            F1--Score         & 0.547 & 0.34 & Avoid (High Drift) \\
            MCC               & 0.641 & 0.40 & Avoid (High Drift) \\
            RES ($\alpha=0.10$) & 0.031 & 0.52 & Recommended (Stable) \\
            RES ($\alpha=0.25$) & 0.077 & 0.38 & Recommended (Stable) \\
            RES ($\alpha=0.50$) & 0.157 & 0.35 & Recommended (Stable) \\
            \hline
        \end{tabular}
    \end{table}

    The link between institutional constraints and $\alpha$ is shown in
    Table~\ref{tab:calibration}. A historical 5\% alarm-rate rule corresponds to
    $\widehat{\alpha}_{A}=0.56$. A 3\% capacity constraint corresponds to
    $\widehat{\alpha}_{B}=0.99$. A loss-based calibration with cost ratio $(1{:}20)$ yields $\widehat{\alpha}_{C}=0.05$ and an induced threshold of approximately 1\%. These values span the full range of operational postures—from highly conservative to highly sensitive-and demonstrate the transparency with which RES metrics encode policy
    preferences.

    \begin{table}[ht]
        \centering
        \caption{Calibrated values of the policy parameter $\alpha$ under three operational
            scenarios.}
        \label{tab:calibration}
        \begin{tabular}{lrr}
            \hline
            Scenario & Calibrated $\alpha$ & Induced Threshold \\
            \hline
            A: Historical (5\%) & 0.56 & 0.05 \\
            B: Capacity (3\%)   & 0.99 & 0.46 \\
            C: Loss (1:20)      & 0.05 & 0.01 \\
            \hline
        \end{tabular}
    \end{table}

    \subsection{Summary}

    The empirical evidence shows that classical metrics are structurally unstable in rare-event credit-default forecasting. Their optimal thresholds drift sharply across prevalence regimes and exhibit substantial volatility within regimes, often yielding operationally implausible alarm rules. RES metrics behave fundamentally differently:
    they maintain interior, stable, and economically interpretable thresholds even under extreme rarity. The calibrated values of $\alpha$ illustrate how the RES framework separates stable institutional preferences from the data-driven implementation threshold $\delta^{\ast}$, enabling consistent decision criteria across models, sampling regimes, and prevalence environments. These results corroborate the theoretical and simulation findings and underline the practical relevance of the RES framework for real-world extreme-risk forecasting applications.

    Supplementary figures, threshold distributions, and additional diagnostics appear in Appendices~G--H (Simulation \& Application diagnostics; full tables in Appendix F).

    \section{Conclusion}

    This paper introduced a class of Rare-Event-Stable (RES) performance metrics designed for binary classification in environments where events occur with extremely low prevalence. We showed, both analytically and empirically, that widely used threshold-dependent metrics such as the F1-score, the Matthews Correlation Coefficient, and Balanced Accuracy possess inherent structural weaknesses in such settings. As the event probability declines, their induced decision thresholds drift systematically across prevalence regimes and exhibit substantial dispersion within fixed regimes, resulting in intervention policies that are operationally inconsistent and difficult to justify.

    The RES framework resolves this instability by introducing a policy parameter $\alpha$ that captures an institution’s relative tolerance for false positives and false negatives. A central feature of the approach is the clear separation between the stable preference parameter $\alpha$ and the data-driven implementation threshold $\delta^{\ast}$. While $\delta^{\ast}$ responds to the empirical score distribution, the institutional preference encoded by $\alpha$ remains invariant across models, samples, and operational environments. This distinction provides a coherent foundation for decision-making in \emph{extreme-imbalance environments}, ensuring that institutional priorities are preserved even as empirical conditions vary.

    The theoretical properties of the RES metrics were validated through extensive simulation experiments spanning five orders of magnitude in prevalence. Classical metrics exhibited the predicted threshold collapse, while RES metrics maintained interior, stable, and economically interpretable decision boundaries throughout. An empirical application using consumer credit-default data reinforced these findings. The RES framework facilitated transparent calibration of $\alpha$ to several operational targets, including historical policy thresholds, institutional capacity constraints, and explicit loss structures. The resulting calibrated values demonstrated how diverse institutional postures are naturally encoded within the RES metric family.

    While the RES framework focuses on decision thresholds, it is not intended to replace proper scoring rules such as the Brier score or the logarithmic score. Proper scoring rules evaluate the calibration and refinement of the entire predictive distribution, providing a necessary foundation for probabilistic forecasting. However, they do not directly guide the binary interventions required in operational settings. We view these approaches as complementary: proper scoring rules should be used to select and calibrate models during development, while RES metrics provide the stability and interpretability required to set operational alarm thresholds in the tail. A holistic evaluation framework for extreme risks should therefore pair a tail-weighted proper scoring rule (for distributional assessment) with an RES metric (for decision-policy definition).

    Future research may extend these ideas to multiclass and multilabel settings, where dependence structures among rare event classes introduce additional challenges. Another promising direction involves reconciling the RES framework with proper scoring rules \citep{gneiting2011comparing}. While RES provides decision-theoretic stability in the extreme tail, proper scoring rules evaluate global probabilistic calibration. Developing a principled way to bridge these complementary objectives remains an open problem in extreme-risk forecasting.

    % ==================================================================
    % REFERENCES
    % ==================================================================
    \bibliography{references}

    \newpage

    \appendix
    \renewcommand{\appendixname}{} % Remove "Appendix"
    \renewcommand{\thesection}{\Alph{section}} % A, B, C numbering
    \renewcommand{\thefigure}{\thesection.\arabic{figure}}
    \renewcommand{\thetable}{\thesection.\arabic{table}}

    \begin{center}
        {\Large \textbf{Appendices}}
    \end{center}

    % ============================================================
    % APPENDIX A: FORMAL DERIVATIONS
    % ============================================================
%    \newpage
    \section{Formal Asymptotic Derivations}
    \label{app:derivations}
    \setcounter{figure}{0}

    This appendix provides the mathematical proofs supporting the structural claims made in Section~5 of the main text regarding the stability of RES metrics and the asymptotic collapse of the F1-score.

    \subsection{Optimality Condition for RES Metrics}

    Recall the definition of the Rare-Event-Stable metric:
    \begin{equation}
        M_{\mathrm{RE}}(\delta; \alpha) = \frac{\mathrm{TPR}(\delta)}{\alpha\,\mathrm{FPR}(\delta) + (1-\alpha)}.
    \end{equation}
    To find the optimal threshold $\delta^{\ast}$, we differentiate $M_{\mathrm{RE}}$ with respect to the decision threshold $\delta$ and set the derivative to zero. Let $f_1(\delta)$ and $f_0(\delta)$ denote the conditional densities of the score for positive and negative cases, respectively. The derivatives of the rates with respect to the threshold are given by $\frac{\partial \mathrm{TPR}}{\partial \delta} = -f_1(\delta)$ and $\frac{\partial \mathrm{FPR}}{\partial \delta} = -f_0(\delta)$.

    Applying the quotient rule for differentiation:
    \begin{equation}
        \frac{\partial M_{\mathrm{RE}}}{\partial \delta}
        =
        \frac{-f_1(\delta)[\alpha\,\mathrm{FPR}(\delta) + (1-\alpha)] - \mathrm{TPR}(\delta)[-\alpha f_0(\delta)]}
        {[\alpha\,\mathrm{FPR}(\delta) + (1-\alpha)]^2} = 0.
    \end{equation}
    For the numerator to be zero, the following equality must hold:
    \[
    f_1(\delta)[\alpha\,\mathrm{FPR}(\delta) + (1-\alpha)] = \alpha\,\mathrm{TPR}(\delta)f_0(\delta).
    \]
    Rearranging terms to isolate the likelihood ratio $\Lambda(\delta) = f_1(\delta)/f_0(\delta)$, we obtain the first-order condition (FOC):
    \begin{equation}
        \Lambda(\delta^{\ast}) = \frac{\alpha\,\mathrm{TPR}(\delta^{\ast})}{\alpha\,\mathrm{FPR}(\delta^{\ast}) + (1-\alpha)}.
        \label{eq:res_foc}
    \end{equation}
    \textbf{Result:} The right-hand side of \eqref{eq:res_foc} contains no prevalence term $\pi$. It depends solely on the shape of the ROC curve (via TPR and FPR) and the policy preference parameter $\alpha$. Consequently, as $\pi \to 0$, the condition defining $\delta^{\ast}$ remains mathematically invariant, ensuring the structural stability of the decision threshold.

    \subsection{Asymptotic Collapse of the F1-Score}

    The F1-score is defined as the harmonic mean of Precision and Recall. Expressed in terms of rates and prevalence $\pi$:
    \begin{equation}
        F1(\delta) = \frac{2\pi\,\mathrm{TPR}(\delta)}{2\pi\,\mathrm{TPR}(\delta) + (1-\pi)\,\mathrm{FPR}(\delta)}.
    \end{equation}
    Maximizing $F1$ is equivalent to maximizing its natural logarithm. Differentiating $\ln(F1)$ with respect to $\delta$ yields:
    \begin{equation}
        \frac{\partial \ln F1}{\partial \delta} = \frac{-f_1(\delta)}{\mathrm{TPR}(\delta)} - \frac{-2\pi f_1(\delta) - (1-\pi)f_0(\delta)}{2\pi\,\mathrm{TPR}(\delta) + (1-\pi)\,\mathrm{FPR}(\delta)} = 0.
    \end{equation}
    Rearranging to solve for the likelihood ratio $\Lambda(\delta^{\ast}) = f_1(\delta^{\ast})/f_0(\delta^{\ast})$ yields:
    \begin{equation}
        \Lambda(\delta^{\ast}) = \left( \frac{1-\pi}{2\pi} \right) \cdot \frac{\mathrm{TPR}(\delta^{\ast})}{2\pi\,\mathrm{TPR}(\delta^{\ast}) + (1-\pi)\,\mathrm{FPR}(\delta^{\ast})}.
        \label{eq:f1_foc}
    \end{equation}
    \textbf{Result:} Consider the limit as $\pi \to 0$. The leading term $\frac{1-\pi}{2\pi}$ diverges to infinity. Unless the second fraction vanishes at a commensurate rate (which implies $\mathrm{TPR} \to 0$), the required likelihood ratio must diverge. For standard unimodal score distributions, a diverging likelihood ratio requirement forces the optimal threshold $\delta^{\ast} \to 1$.

    \begin{figure}[ht]
        \centering
        \includegraphics[width=0.85\linewidth]{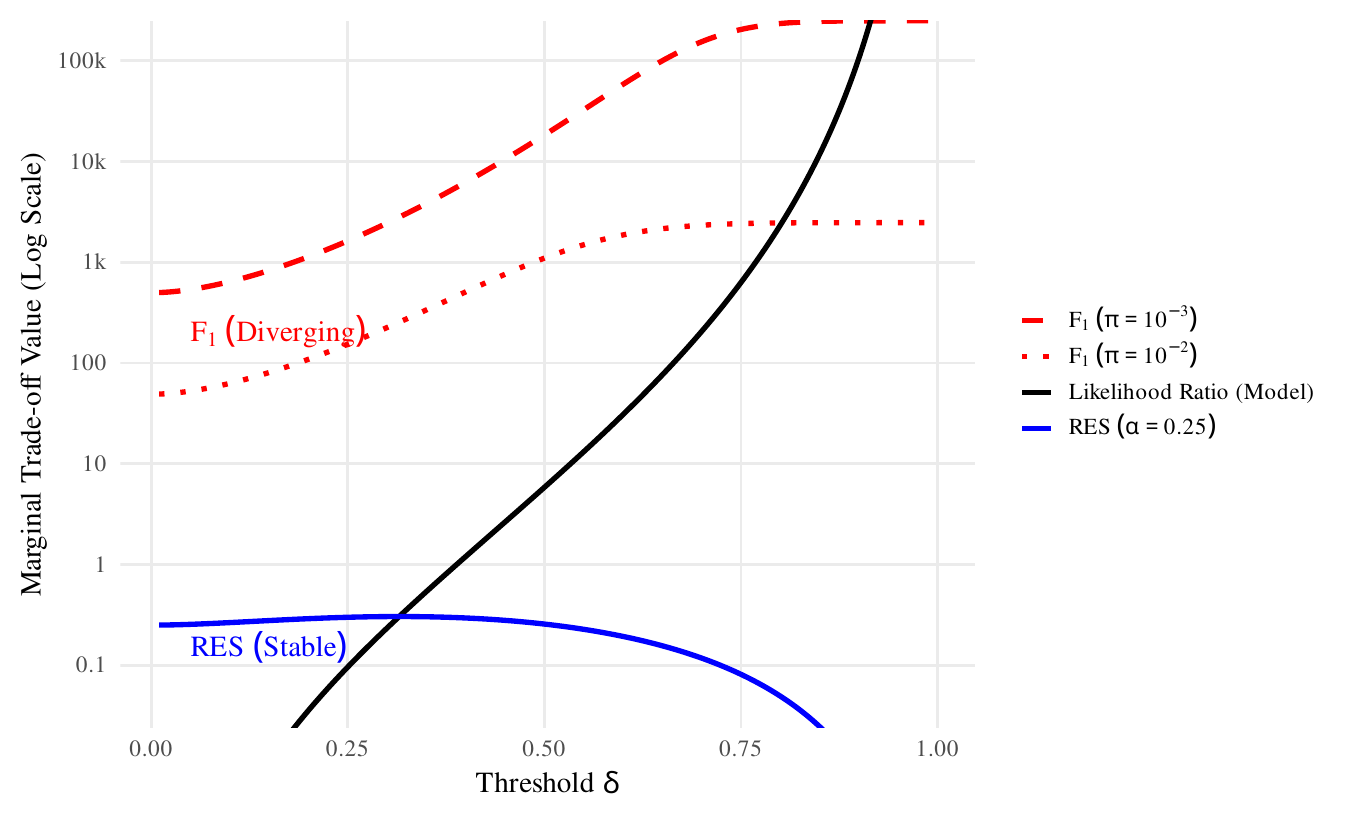}
        \caption{\textbf{Geometric mechanism of threshold collapse.} The optimal threshold $\delta^\ast$ is determined by the intersection of the model likelihood ratio and the metric’s marginal trade-off curve. The \textcolor{blue}{RES curve} depends only on $\alpha$ and stays fixed, yielding a stable interior intersection consistent with \Cref{eq:res_foc}. In contrast, the \textcolor{red}{$F_1$ curves} diverge upward with decreasing prevalence ($10^{-2} \to 10^{-3}$), pushing $\delta^\ast \to 1$ as implied by \Cref{eq:f1_foc}. Note the logarithmic scale on the y-axis.}
        \label{fig:analytical_mechanism}
    \end{figure}

    % ============================================================
    % APPENDIX B: COMPUTATIONAL STRATEGY (NEW)
    % ============================================================
    \newpage
    \section{Computational Strategy for Extreme Rarity}
    \label{app:computational}
    \setcounter{figure}{0}

    Simulating rare events at prevalence levels as low as $\pi = 10^{-6}$ presents significant computational challenges. With a fixed target of $n_{+} = 20$ positive events per replication, the implied number of negative observations required to maintain the prevalence ratio is approximately $20$ million. Storing, sorting, and processing vectors of this size for thousands of Monte Carlo replications creates memory bottlenecks that can impede efficient analysis.

    To address this, the simulation framework employs a \emph{Weighted Downsampling Strategy} that preserves the statistical properties of the full sample while reducing memory usage by approximately 90\%.

    \subsection{Weighted Data Generation}

    Let $N_{\mathrm{req}} = n_{+}(1-\pi)/\pi$ be the required number of negative observations to achieve the target prevalence $\pi$. We define a computational capacity constraint $N_{\mathrm{cap}} = 2 \times 10^6$. The actual number of simulated negatives $n_{-}$ and their associated weights $w_{-}$ are determined as follows:

    \begin{itemize}
        \item \textbf{Positives:} Simulate $n_{+}$ observations from the conditional distribution $F_1$. Assign each a weight $w_{+} = 1$.
        \item \textbf{Negatives:}
        \begin{itemize}
            \item If $N_{\mathrm{req}} \le N_{\mathrm{cap}}$, simulate $N_{\mathrm{req}}$ observations from $F_0$. Assign each a weight $w_{-} = 1$.
            \item If $N_{\mathrm{req}} > N_{\mathrm{cap}}$, simulate $N_{\mathrm{cap}}$ observations from $F_0$. Assign each a weight $w_{-} = \frac{N_{\mathrm{req}}}{N_{\mathrm{cap}}}$.
        \end{itemize}
    \end{itemize}

    \subsection{Metric Calculation via Weighted Counts}

    Confusion matrix components are calculated using the weighted cumulative sums of the observations sorted by score. Let $y_{(i)}$ be the binary label and $w_{(i)}$ be the weight of the $i$-th observation in the sorted sequence. The weighted True Positives ($TP_w$) and False Positives ($FP_w$) at rank $k$ are calculated as:
    \[
    TP_w(k) = \sum_{i=1}^{k} \mathbb{I}(y_{(i)}=1) \cdot w_{(i)}, \quad
    FP_w(k) = \sum_{i=1}^{k} \mathbb{I}(y_{(i)}=0) \cdot w_{(i)}.
    \]
    The TPR and FPR are then computed as $TP_w(k) / \sum w_{+}$ and $FP_w(k) / \sum w_{-}$, respectively. Because the downsampling is uniform within the negative class, $FP_w$ is an unbiased estimator of the false positive count in the full population. This approach ensures that the resulting ROC curves, thresholds, and metric evaluations are statistically indistinguishable from the full-sample case while fitting within standard RAM constraints.

    % ============================================================
    % APPENDIX C: SIMULATION DETAILS (ORIGINAL FULL TEXT)
    % ============================================================
    \newpage
    \section{Simulation Details}
    \label{app:simulation}
    \setcounter{figure}{0}

    This appendix documents the simulation framework supporting Section~6 of the paper.
    The objective is to evaluate the behaviour of traditional threshold-based metrics and
    Rare-Event-Stable (RES) metrics across orders of magnitude variation in event
    prevalence. The design ensures a controlled environment in which any instability or
    collapse of a metric can be attributed to its structural properties rather than to
    sampling noise or model misspecification.

    \subsection{Data-Generating Process}

    Each simulation draws observations from fully specified conditional forecast
    distributions:
    \[
    \eta(X)\mid Y=1 \sim F_1,
    \qquad
    \eta(X)\mid Y=0 \sim F_0,
    \]
    where $\eta(X)\in(0,1)$ is the model-implied score.

    Two regimes of tail discrimination are considered:

    \begin{itemize}
        \item \textbf{Moderate separation:}
        \[
        F_1 = \mathrm{Beta}(5,3),
        \qquad
        F_0 = \mathrm{Beta}(2,8).
        \]

        \item \textbf{Strong separation:}
        \[
        F_1 = \mathrm{Beta}(8,2),
        \qquad
        F_0 = \mathrm{Beta}(1,12).
        \]
    \end{itemize}

    These distributions satisfy the monotone-likelihood-ratio condition and permit precise
    control over the degree of overlap between the positive and negative classes.

    \subsection{Prevalence Levels and Sample Sizes}

    Event prevalence varies over five orders of magnitude:
    \[
    \pi \in \{10^{-2},\,10^{-3},\,10^{-4},\,10^{-5},\,10^{-6}\}.
    \]

    To isolate prevalence effects while avoiding resampling noise:

    \begin{itemize}
        \item For $\pi \ge 10^{-5}$, each replication includes $n_{+}=100$ positive cases.
        \item For $\pi \le 10^{-5}$, each replication includes $n_{+}=20$ positive cases.
    \end{itemize}

    Given $n_{+}$, the number of negative observations is
    \[
    n_{-}(\pi)
    =
    \left\lfloor n_{+} \,\frac{1-\pi}{\pi}\right\rceil,
    \]
    so that total sample size scales approximately as $1/\pi$, matching the geometry of
    rare-event environments encountered in credit risk, fraud detection and reliability
    engineering.

    \subsection{Monte Carlo Structure}

    For each pair of (signal regime, prevalence), the experiment performs
    \[
    R = 2{,}000
    \]
    independent Monte Carlo replications.
    For each replication:

    \begin{enumerate}
        \item Draw $n_{+}$ samples from $F_1$ and $n_{-}$ samples from $F_0$.
        \item Sort all scores in decreasing order.
        \item Construct the full confusion-matrix path by evaluating $(\TPR(\delta),\FPR(\delta))$
        at each distinct threshold.
        \item Evaluate all metrics along this path.
        \item Record the optimal threshold $\delta^{\ast}$ for each metric.
    \end{enumerate}

    This produces full empirical distributions of optimal thresholds and metric values at
    each prevalence level.

    \subsection{Metrics Evaluated}

    The simulation compares the following metrics:

    \begin{itemize}
        \item \textbf{F1-score}
        \[
        F1(\delta)
        =
        \frac{2 \,\mathrm{Prec}(\delta)\,\TPR(\delta)}
        {\mathrm{Prec}(\delta)+\TPR(\delta)}.
        \]

        \item \textbf{Balanced Accuracy}
        \[
        \mathrm{BA}(\delta)
        =
        \tfrac{1}{2}\left(\TPR(\delta)+\TNR(\delta)\right).
        \]

        \item \textbf{AUC (ROC)}
        Computed when $N<10^6$; omitted otherwise for computational efficiency.

        \item \textbf{RES metric} for selected preference levels:
        \[
        M_{\mathrm{RE}}(\delta;\alpha)
        =
        \frac{\TPR(\delta)}
        {\alpha\,\FPR(\delta)+(1-\alpha)},
        \qquad
        \alpha \in\{0.10,\,0.50\}.
        \]
    \end{itemize}

    Optimal thresholds are defined as
    \[
    \delta^{\ast}
    =
    \arg\max_{\delta} M(\delta),
    \]
    with ties broken in favour of the smallest threshold.

    % ============================================================
    % APPENDIX D: APPLICATION (FULL TEXT + HYPERPARAMS)
    % ============================================================
    \newpage
    \section{Application: Credit-Default Forecasting}
    \label{app:application}
    \setcounter{figure}{0}
    \setcounter{table}{0}

    This appendix describes the empirical environment used in Section~7. The objective is
    to evaluate the behaviour of classical and Rare-Event-Stable (RES) metrics in a real
    credit-risk setting, where default events are rare and forecast distributions exhibit
    substantial upper-tail concentration.

    \subsection{Data}

    The analysis relies on the \textit{Give Me Some Credit} consumer-credit dataset,
    a widely used benchmark containing borrower characteristics and a binary indicator of
    serious delinquency within two years. After removing non-informative fields and
    applying median imputation to missing numeric entries, the resulting dataset contains
    only predictive covariates and the binary outcome.

    A stratified split allocates 70\% of observations to the training set and 30\% to the
    test set while preserving the empirical default rate ($\approx 6.7\%$). All metric
    evaluations and calibrations use out-of-sample predicted probabilities from the test
    set.

    \subsection{Forecasting Model}

    Probability-of-default forecasts are generated using a LightGBM model. Table~\ref{tab:hyperparameters} lists the exact hyperparameters used in the analysis. These values were selected to ensure a stable, monotonic score distribution typical of credit risk models, avoiding aggressive overfitting.

    \begin{table}[ht]
        \centering
        \caption{LightGBM Hyperparameters used in the empirical forecasting model.}
        \label{tab:hyperparameters}
        \small
        \begin{tabular}{lll}
            \toprule
            \textbf{Parameter} & \textbf{Value} & \textbf{Justification} \\
            \midrule
            \texttt{objective} & \texttt{binary} & Standard Log-Loss minimization \\
            \texttt{learning\_rate} & 0.05 & Conservative learning rate for stability \\
            \texttt{num\_leaves} & 31 & Limits interactions to prevent overfitting \\
            \texttt{bagging\_fraction} & 0.8 & Stochastic subsampling for robustness \\
            \texttt{bagging\_freq} & 1 & Frequency of bagging (every iteration) \\
            \texttt{nrounds} & 450 & Determined via early stopping analysis \\
            \bottomrule
        \end{tabular}
    \end{table}

    Denote the test-set probability forecasts by $\widehat{p}_i$ and the corresponding
    default indicators by $Y_i$.

%    \subsection{Forecasting Model}
%
%    Probability-of-default forecasts are generated using a LightGBM model trained in \textsf{R} via the \texttt{lightgbm} package \citep{ke2017lightgbm}. The following hyperparameters are used:
%
%    \begin{itemize}
%        \setlength\itemsep{0em} % Makes the list more compact
%        \item \texttt{objective = "binary"};
%        \item \texttt{metric = "binary\_logloss"};
%        \item \texttt{learning\_rate = 0.05};
%        \item \texttt{num\_leaves = 31};
%        \item \texttt{min\_data\_in\_leaf = 40};
%        \item \texttt{feature\_fraction = 0.9};
%        \item \texttt{bagging\_fraction = 0.8};
%        \item \texttt{bagging\_freq = 1};
%        \item \texttt{nrounds = 450}.
%    \end{itemize}
%
%    These settings produce numerically stable and reasonably calibrated probability forecasts while avoiding overfitting and requiring minimal tuning. The purpose of the empirical experiment is not to identify the best-performing predictive model, but rather to provide a consistent forecast distribution for evaluating the behaviour of classical and RES metrics under operationally relevant default rates. Consequently, extensive hyperparameter search, Bayesian optimisation, and model comparisons are deliberately avoided.

    \subsection{Controlled Prevalence Regimes}

    To study performance under increasing rarity, additional evaluation regimes are
    constructed by down-sampling positive test-set cases while retaining all negatives.
    Given a target prevalence $\pi_{\mathrm{target}}$, the required number of positive
    observations is
    \[
    n_{+}
    =
    \frac{\pi_{\mathrm{target}}}{1-\pi_{\mathrm{target}}}\,n_{-}.
    \]

    If $n_{+}$ exceeds the available positive observations, the regime defaults to the full
    test set. The analysis uses the following target prevalences:
    \[
    \{0.001,\;0.005,\;0.010,\;0.020,\;\text{empirical prevalence}\}.
    \]

    This design generates realistic low-prevalence environments while preserving the score
    distribution of the forecasting model.

    \subsection{Bootstrap Framework}

    Within each prevalence regime, $500$ bootstrap samples of fixed size are drawn. For
    each bootstrap replication:

    \begin{enumerate}
        \item Sort predicted probabilities $\widehat{p}_i$ in descending order.
        \item Construct the confusion-matrix path $(\TPR(\delta),\FPR(\delta))$ along all
        distinct thresholds.
        \item Evaluate each metric at each threshold.
        \item Record the optimal threshold $\delta^{\ast}$ for each metric.
    \end{enumerate}

    This produces empirical distributions of both metric values and optimal thresholds
    across prevalence levels, enabling direct comparison between methods and identifying
    threshold instability or collapse.

    % ============================================================
    % APPENDIX E: CALIBRATION ALGORITHMS (FULL)
    % ============================================================
    \newpage
    \section{Calibration of the Policy Parameter $\alpha$}
    \label{app:calibration}
    \setcounter{figure}{0}

    This appendix describes general procedures for calibrating the RES policy parameter
    $\alpha$. While the implementation threshold $\delta^{\ast}$ adapts to the empirical
    forecast distribution, $\alpha$ encodes a stable institutional preference concerning
    the marginal trade-off between false positives and false negatives.

    \subsection{Cost-Based Calibration}

    If false-positive and false-negative errors have explicit institutional costs
    $C_{\mathrm{FP}}$ and $C_{\mathrm{FN}}$, the RES marginal trade-off implies:
    \[
    \frac{\alpha}{1-\alpha}
    =
    \frac{C_{\mathrm{FP}}}{C_{\mathrm{FN}}}
    \quad\Longrightarrow\quad
    \alpha
    =
    \frac{C_{\mathrm{FP}}}{C_{\mathrm{FP}}+C_{\mathrm{FN}}}.
    \]

    \begin{algorithm}[H]
        \caption{Cost-Based Calibration of $\alpha$}
        \begin{algorithmic}
            \Require Misclassification costs $(C_{\mathrm{FP}}, C_{\mathrm{FN}})$
            \State Compute $r = C_{\mathrm{FP}}/(C_{\mathrm{FP}} + C_{\mathrm{FN}})$
            \State \Return $\alpha^{\ast} = r$
        \end{algorithmic}
    \end{algorithm}

    \subsection{Calibration to a Historical Threshold}

    Institutions often use legacy probability-of-default cutoffs such as
    $\delta_{\mathrm{hist}} = 0.05$.
    Calibration selects $\alpha$ such that the induced RES-optimal threshold satisfies
    $\delta^{\ast}_{\mathrm{RE}}(\alpha) \approx \delta_{\mathrm{hist}}$.

    \begin{algorithm}[H]
        \caption{Calibration to a Historical Threshold}
        \begin{algorithmic}
            \Require Scores $\eta_i$, labels $Y_i$, target threshold $\delta_{\mathrm{hist}}$
            \State Construct confusion-matrix path $(\TPR(\delta),\FPR(\delta))$
            \For{$\alpha$ in grid $\mathcal{A}$}
            \State Compute $M_{\mathrm{RE}}(\delta;\alpha)$ along the path
            \State Set $\delta^{\ast}(\alpha) = \arg\max_{\delta} M_{\mathrm{RE}}(\delta;\alpha)$
            \EndFor
            \State \Return $\alpha^{\ast} = \arg\min_{\alpha} |\delta^{\ast}(\alpha)-\delta_{\mathrm{hist}}|$
        \end{algorithmic}
    \end{algorithm}

    \subsection{Calibration to an Intervention Rate}

    If the institution can process at most
    $r_{\mathrm{target}}$ fraction of alerts, calibration requires matching the induced alarm rate
    $r(\alpha) = \Pr(\eta_i \ge \delta^{\ast}_{\mathrm{RE}}(\alpha))$.

    \begin{algorithm}[H]
        \caption{Calibration to an Intervention-Rate Constraint}
        \begin{algorithmic}
            \Require Scores $\eta_i$, labels $Y_i$, intervention target $r_{\mathrm{target}}$
            \State Construct confusion-matrix path $(\TPR(\delta),\FPR(\delta))$
            \For{$\alpha$ in grid $\mathcal{A}$}
            \State Compute $\delta^{\ast}(\alpha)$
            \State Compute $r(\alpha) = \Pr(\eta_i \ge \delta^{\ast}(\alpha))$
            \EndFor
            \State \Return $\alpha^{\ast} = \arg\min_{\alpha} |r(\alpha)-r_{\mathrm{target}}|$
        \end{algorithmic}
    \end{algorithm}

    \subsection{Loss-Based Calibration}

    For an institutional loss function
    $L(\delta) = C_{\mathrm{FN}} (1-\TPR(\delta)) + C_{\mathrm{FP}}\,\FPR(\delta)$,
    the optimal decision threshold is $\delta^{\ast}_{\mathrm{loss}} = \arg\min_{\delta} L(\delta)$.
    Calibration selects $\alpha$ so that the RES-optimal threshold matches this loss-optimal threshold.

    \begin{algorithm}[H]
        \caption{Loss-Based Calibration of $\alpha$}
        \begin{algorithmic}
            \Require Scores $\eta_i$, labels $Y_i$, misclassification costs $(C_{\mathrm{FP}},C_{\mathrm{FN}})$
            \State Compute loss-optimal threshold $\delta^{\ast}_{\mathrm{loss}}$
            \For{$\alpha$ \textbf{in} grid $\mathcal{A}$}
            \State Compute $\delta^{\ast}(\alpha)$
            \EndFor
            \State \Return $\alpha^{\ast} = \arg\min_{\alpha} |\delta^{\ast}(\alpha) - \delta^{\ast}_{\mathrm{loss}}|$
        \end{algorithmic}
    \end{algorithm}

    % ============================================================
    % APPENDIX F: EMPIRICAL RESULTS (FULL TABLES)
    % ============================================================
    \newpage
    \section{Empirical Calibration and Results}
    \label{app:empirical_calibration}
    \setcounter{figure}{0}
    \setcounter{table}{0}

    This appendix presents the full numerical results from the credit-default application in Section~7.

    \subsection{Calibrated Values}

    Applying the three selected calibration methods to the LightGBM probability forecasts yields the following estimates (see Table~\ref{tab:calibration}).

    \begin{table}[ht]
        \centering
        \caption{Calibrated values of $\alpha$ and induced thresholds.}
        \label{tab:calibration}
        \begin{tabular}{lrr}
            \toprule
            Scenario & Calibrated $\alpha$ & Induced Threshold \\
            \midrule
            A: Historical (5\%) & 0.56 & 0.05 \\
            B: Capacity (3\%)   & 0.99 & 0.46 \\
            C: Loss (1:20)      & 0.05 & 0.01 \\
            \bottomrule
        \end{tabular}
    \end{table}

    \begin{itemize}
        \item \textbf{Historical-threshold calibration:} $\widehat{\alpha}_{A} = 0.56$.
        \item \textbf{Intervention-rate calibration:} $\widehat{\alpha}_{B} = 0.99$.
        \item \textbf{Loss-based calibration:} $\widehat{\alpha}_{C} = 0.05$.
    \end{itemize}

    The empirically calibrated value $\widehat{\alpha}_{C} = 0.05$ aligns closely with the theoretical optimum derived from the cost ratio $1:20$, which implies a theoretical $\alpha = 1/(1+20) \approx 0.048$. This confirms that the RES metric correctly internalizes the provided asymmetric cost structure.

    \subsection{Prevalence Regimes}
    The following table details the exact sample sizes used for the down-sampling regimes.

    \begin{table}[ht]
        \centering
        \caption{Prevalence regimes used for the empirical evaluation.}
        \label{tab:B1}
        \begin{tabular}{rrrrr}
            \toprule
            Target $\pi$ & Empirical $\pi$ & $N_{\text{total}}$ & $N_{+}$ & $N_{-}$ \\
            \midrule
            0.0010 & 0.0010 & 42056 &  42  & 42014 \\
            0.0050 & 0.0050 & 42225 & 211  & 42014 \\
            0.0100 & 0.0100 & 42438 & 424  & 42014 \\
            0.0200 & 0.0200 & 42871 & 857  & 42014 \\
            0.0664 & 0.0664 & 45000 & 2986 & 42014 \\
            \bottomrule
        \end{tabular}
    \end{table}

    \subsection{Stability Analysis Tables}
    Tables \ref{tab:C1} and \ref{tab:C2} report the stability metrics for Classical and RES approaches, respectively.

    \begin{table}[ht]
        \centering
        \caption{Classical metrics: threshold stability across regimes.}
        \label{tab:C1}
        \begin{tabular}{lrrrr}
            \toprule
            Metric & Min & Max & Range & CV \\
            \midrule
            Balanced Accuracy & 0.02 & 0.19 & 0.17 & 0.26 \\
            F1-Score          & 0.16 & 0.71 & 0.55 & 0.34 \\
            MCC               & 0.07 & 0.71 & 0.64 & 0.40 \\
            \bottomrule
        \end{tabular}
    \end{table}

    \begin{table}[ht]
        \centering
        \caption{RES metrics: stability profiles across regimes.}
        \label{tab:C2}
        \begin{tabular}{rrrr}
            \toprule
            $\alpha$ & Mean Threshold & Threshold CV & MRE CV \\
            \midrule
            0.10 & 0.01 & 0.52 & 0.01 \\
            0.25 & 0.02 & 0.38 & 0.02 \\
            0.50 & 0.04 & 0.35 & 0.05 \\
            \bottomrule
        \end{tabular}
    \end{table}

    %    \begin{table}[ht]
        %        \centering
        %        \caption{Operational guidance under extreme imbalance.}
        %        \label{tab:E1}
        %        \begin{tabular}{lrrl}
            %            \toprule
            %            Metric & Range & CV & Recommendation \\
            %            \midrule
            %            Balanced Accuracy  & 0.174 & 0.26 & Symmetric Only \\
            %            F1-Score           & 0.547 & 0.34 & Avoid (High Drift) \\
            %            MCC                & 0.641 & 0.40 & Avoid (High Drift) \\
            %            RES ($\alpha{=}0.10$) & 0.031 & 0.52 & Recommended (Stable) \\
            %            RES ($\alpha{=}0.25$) & 0.077 & 0.38 & Recommended (Stable) \\
            %            RES ($\alpha{=}0.50$) & 0.157 & 0.35 & Recommended (Stable) \\
            %            \bottomrule
            %        \end{tabular}
        %    \end{table}
    %
    \begin{table}[ht]
        \centering
        \begin{threeparttable}
            \caption{Operational guidance under extreme imbalance.}
            \label{tab:E1_appendix}
            \begin{tabular}{lrrl}
                \toprule
                Metric & Range & CV & Recommendation \\
                \midrule
                Balanced Accuracy  & 0.174 & 0.26 & Symmetric Only \\
                F1-Score           & 0.547 & 0.34 & Avoid (High Drift) \\
                MCC                & 0.641 & 0.40 & Avoid (High Drift) \\
                RES ($\alpha{=}0.10$) & 0.031 & 0.52 & Recommended (Stable) \\
                RES ($\alpha{=}0.25$) & 0.077 & 0.38 & Recommended (Stable) \\
                RES ($\alpha{=}0.50$) & 0.157 & 0.35 & Recommended (Stable) \\
                \bottomrule
            \end{tabular}
            \begin{tablenotes}\footnotesize
                \item \textit{Note:} A higher CV for RES $\delta^{\ast}$ can remain operationally acceptable because the absolute thresholds stay within a narrow interior band (e.g., $0.01$–$0.05$). Variability at this scale does not materially alter review volumes, unlike classical metrics whose thresholds drift toward degenerate extremes.
            \end{tablenotes}
        \end{threeparttable}
    \end{table}

    % ============================================================
    % APPENDIX G: SIMULATION FIGURES (EMBEDDED)
    % ============================================================
    \newpage
    \section{Additional Figures: Simulation Diagnostics}
    \label{app:sim_figures}
    \setcounter{figure}{0}

    This appendix contains the supplementary diagnostic figures associated with the Simulation Study (Section 6).

    \begin{figure}[H]
        \centering
        \includegraphics[width=0.75\linewidth]{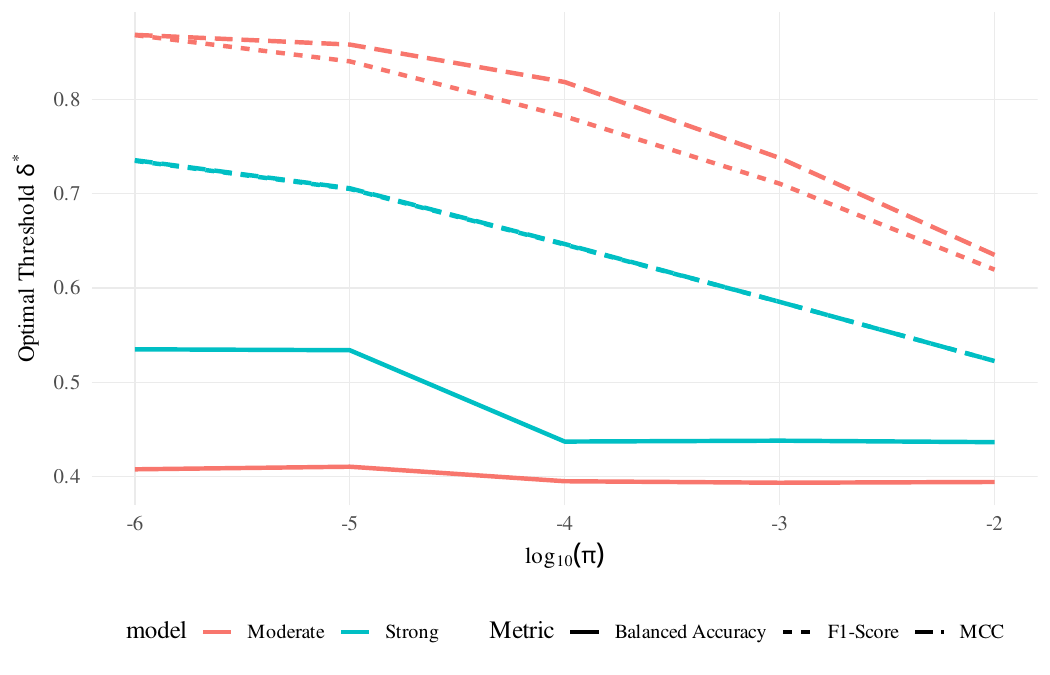}
        \caption{\textbf{Simulation Series A1: Classical Threshold Collapse.} Comparison of threshold drift for F1-score, MCC, and Balanced Accuracy across five orders of magnitude in prevalence.}
        \label{fig:sim_A1}
    \end{figure}

    \begin{figure}[H]
        \centering
        \includegraphics[width=0.75\linewidth]{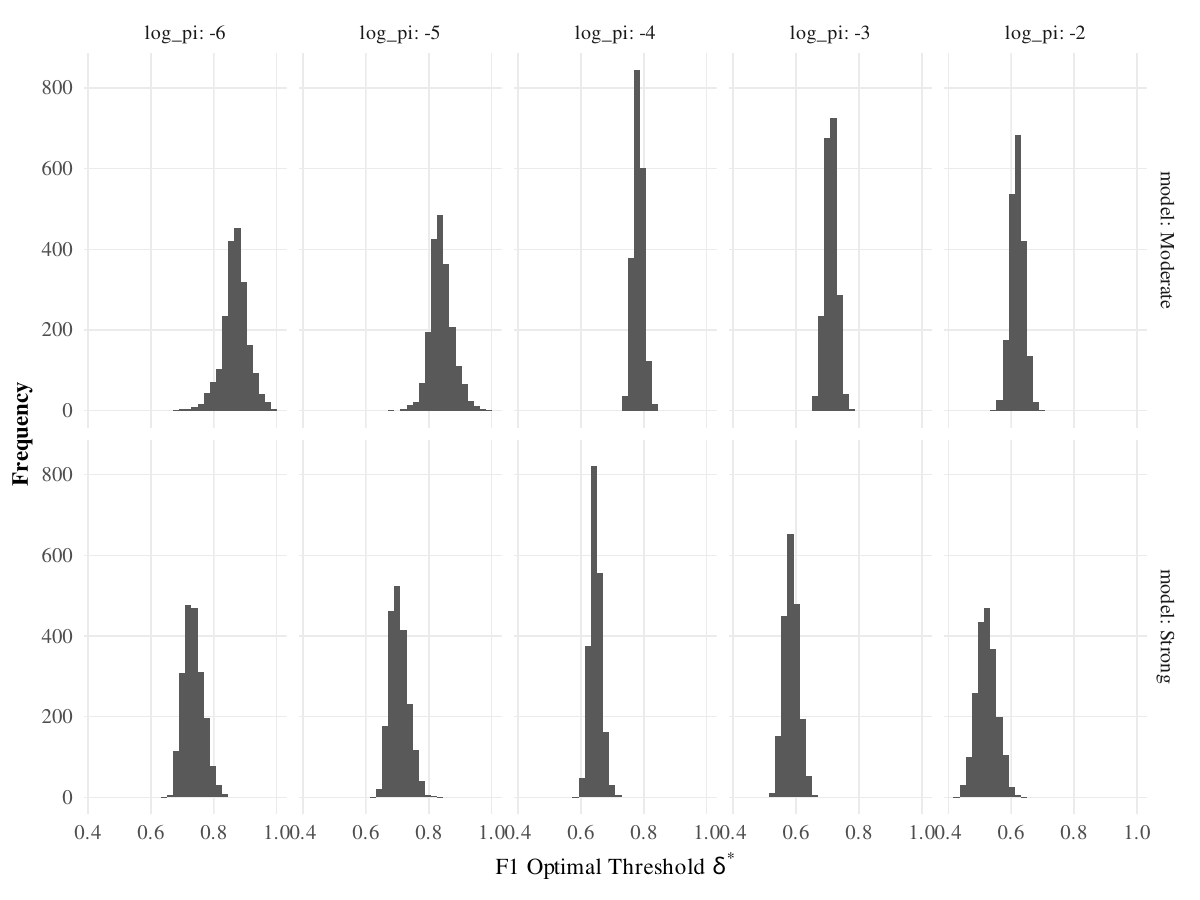}
        \caption{\textbf{Simulation Series A2: F1 Threshold Distributions.} Bootstrap histograms of optimal F1 thresholds showing increasing dispersion and multi-modality as prevalence declines.}
        \label{fig:sim_A2}
    \end{figure}

    \begin{figure}[H]
        \centering
        \includegraphics[width=0.75\linewidth]{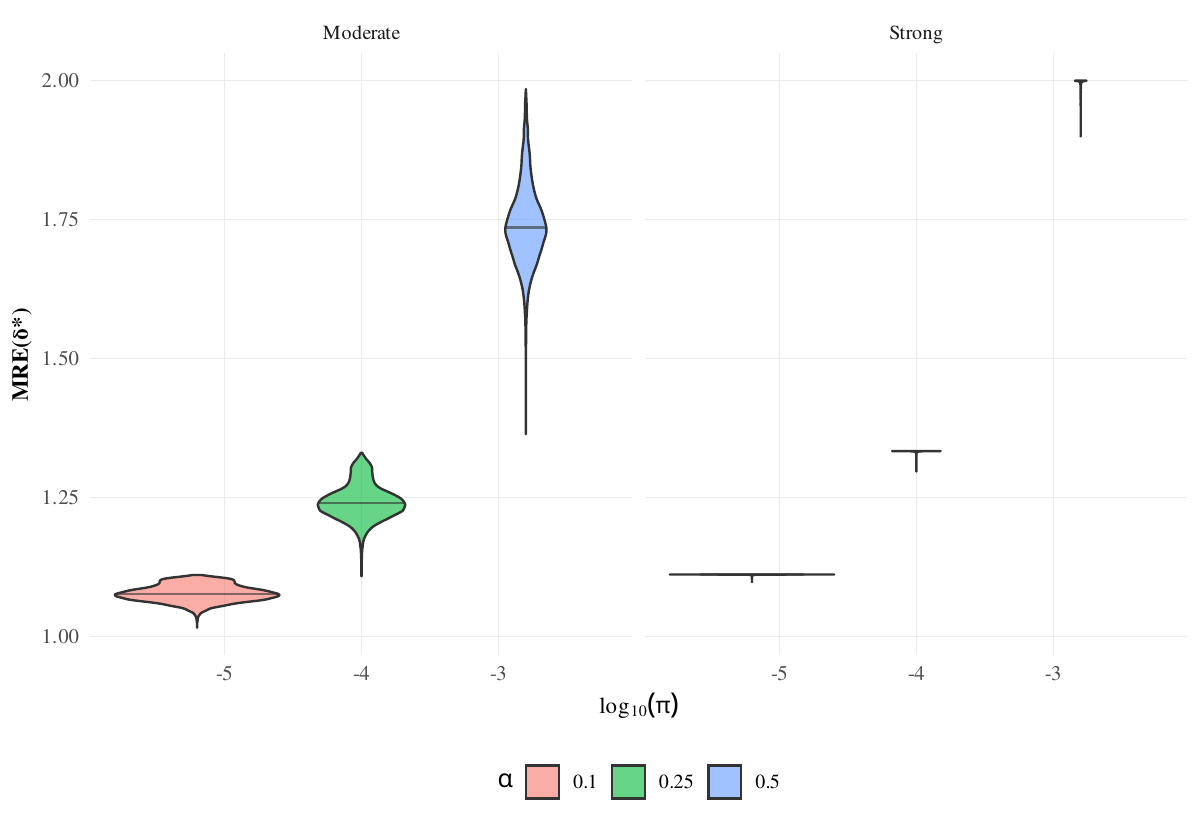}
        \caption{\textbf{Simulation Series A4-a: RES Metric Value Stability.} Violin plots of the maximised RES metric value $M(\delta^{\ast})$ across prevalence regimes. Unlike classical metrics which decay or saturate, RES values remain strictly bounded and distinct for each $\alpha$, preserving discriminative signal strength even at $\pi=10^{-6}$.}
        \label{fig:sim_A4a}
    \end{figure}

    \begin{figure}[H]
        \centering
        \includegraphics[width=0.75\linewidth]{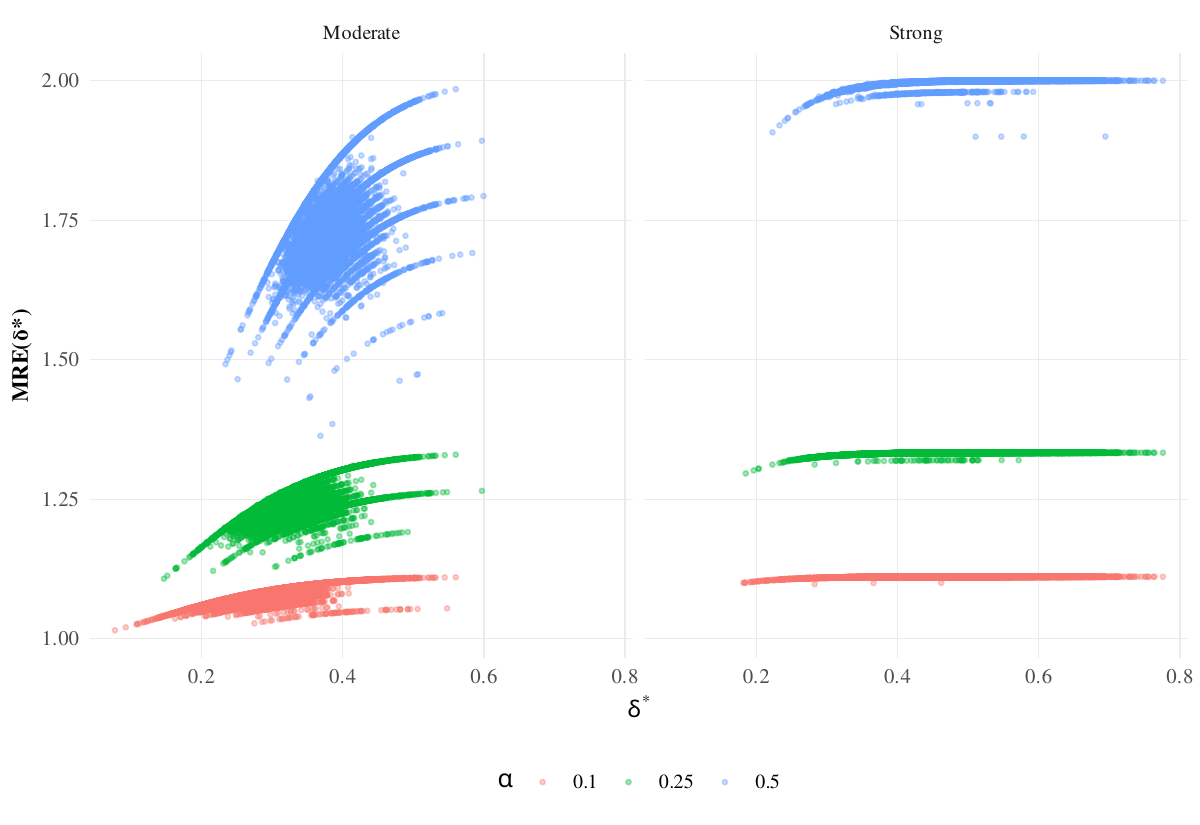}
        \caption{\textbf{Simulation Series A4-b: Geometry of RES Optimisation.} Scatter plots of optimal threshold $\delta^{\ast}$ vs. maximum metric value. The data reveal strictly stratified operating bands, demonstrating that $\alpha$ effectively separates institutional preferences from stochastic sampling variation.}
        \label{fig:sim_A4b}
    \end{figure}

    \begin{figure}[H]
        \centering
        \includegraphics[width=0.75\linewidth]{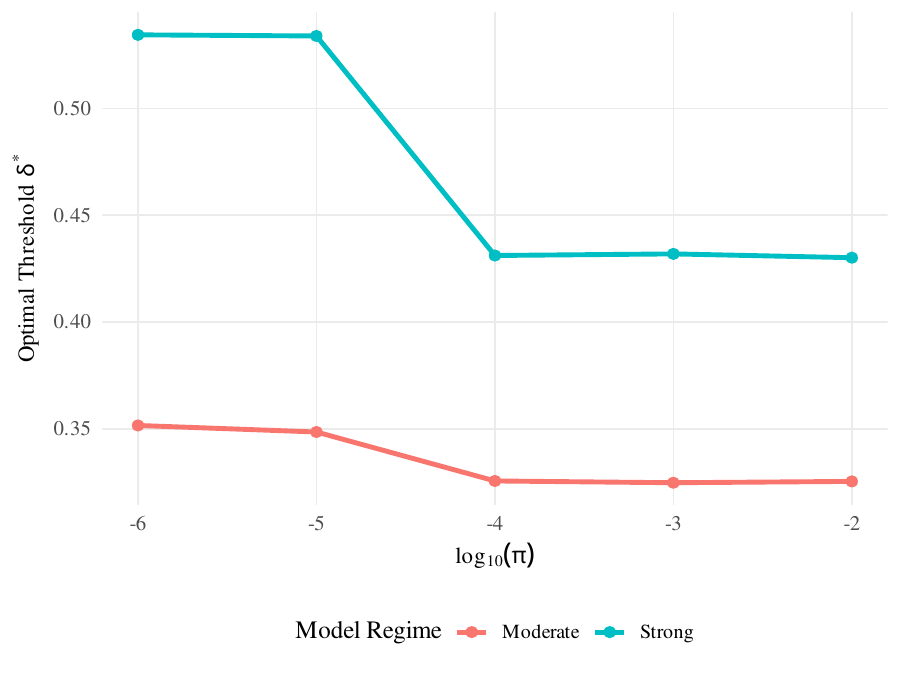}
        \caption{\textbf{Simulation Series A3: RES Threshold Paths ($\alpha=0.25$).} Optimal RES thresholds remain strictly interior and stable across all prevalence regimes.}
        \label{fig:sim_A3}
    \end{figure}

    \begin{figure}[H]
        \centering
        \includegraphics[width=0.75\linewidth]{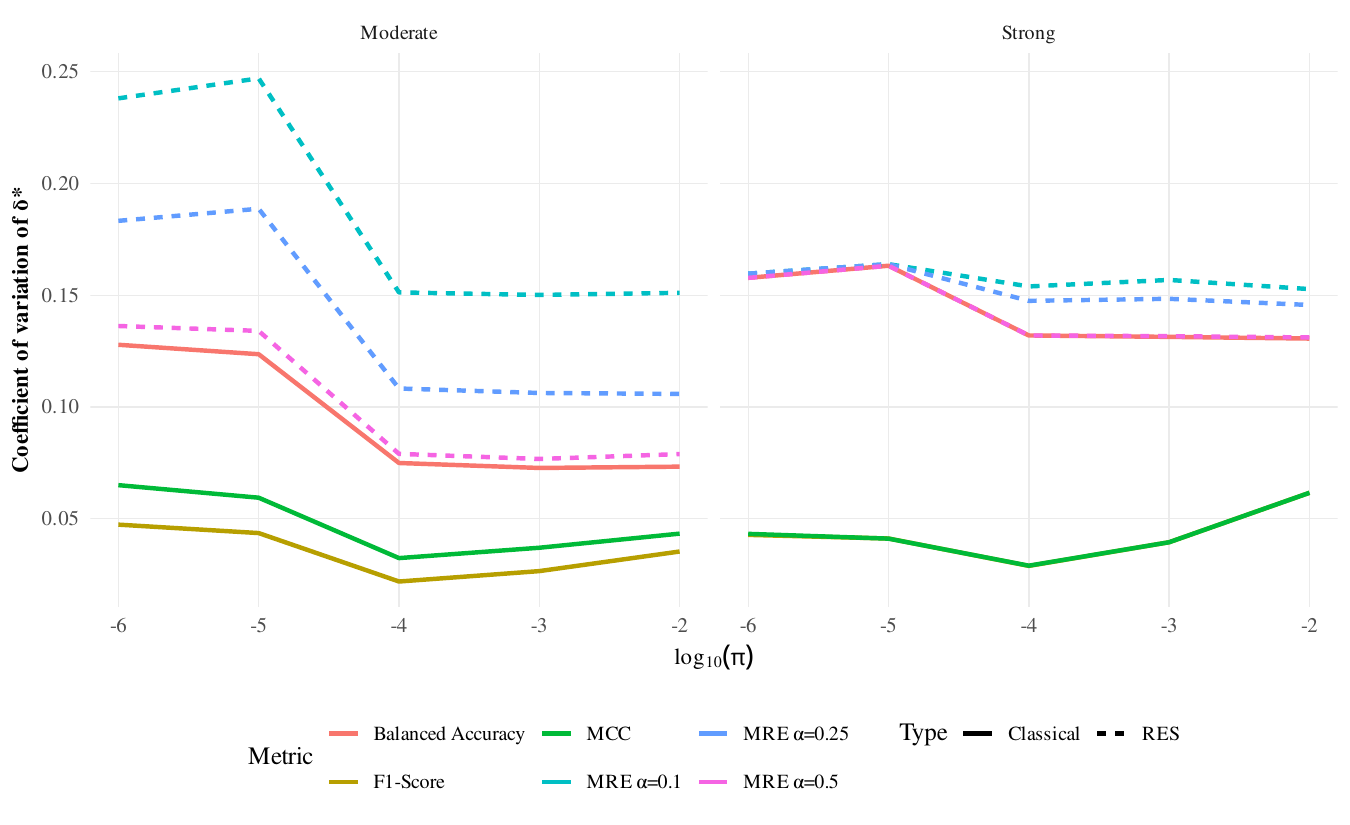}
        \caption{\textbf{Simulation Series A4: Stability Diagnostics.} Coefficient of Variation (CV) of the optimal threshold for Classical vs RES metrics.}
        \label{fig:sim_A4}
    \end{figure}

    \begin{figure}[H]
        \centering
        \includegraphics[width=0.75\linewidth]{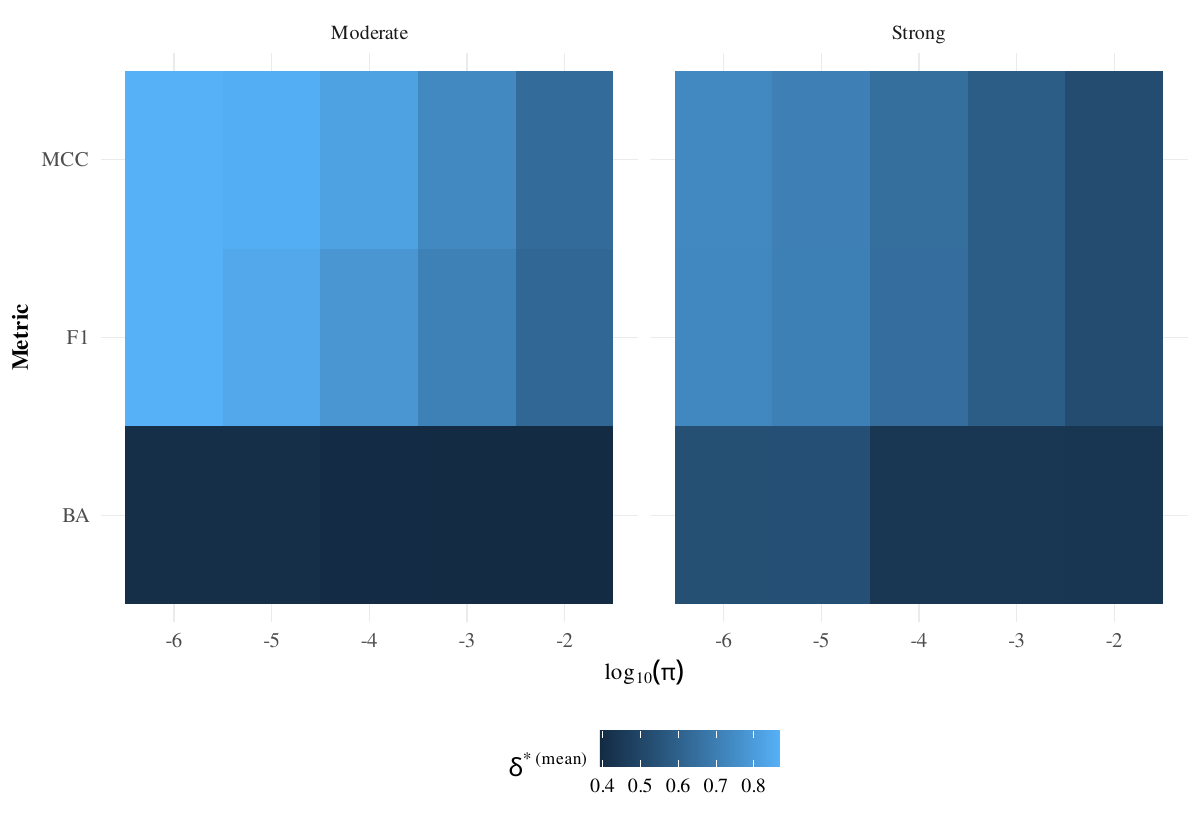}
        \caption{\textbf{Simulation Series A5: Threshold Heatmap.} Evolution of the mean optimal threshold as a function of metric and prevalence.}
        \label{fig:sim_A5}
    \end{figure}

    % ============================================================
    % APPENDIX H: APPLICATION FIGURES (EMBEDDED)
    % ============================================================
    \newpage
    \section{Additional Figures: Application Diagnostics}
    \label{app:app_figures}
    \setcounter{figure}{0}

    This appendix contains the supplementary diagnostic figures associated with the Empirical Application (Section 7).

    \begin{figure}[H]
        \centering
        \includegraphics[width=0.75\linewidth]{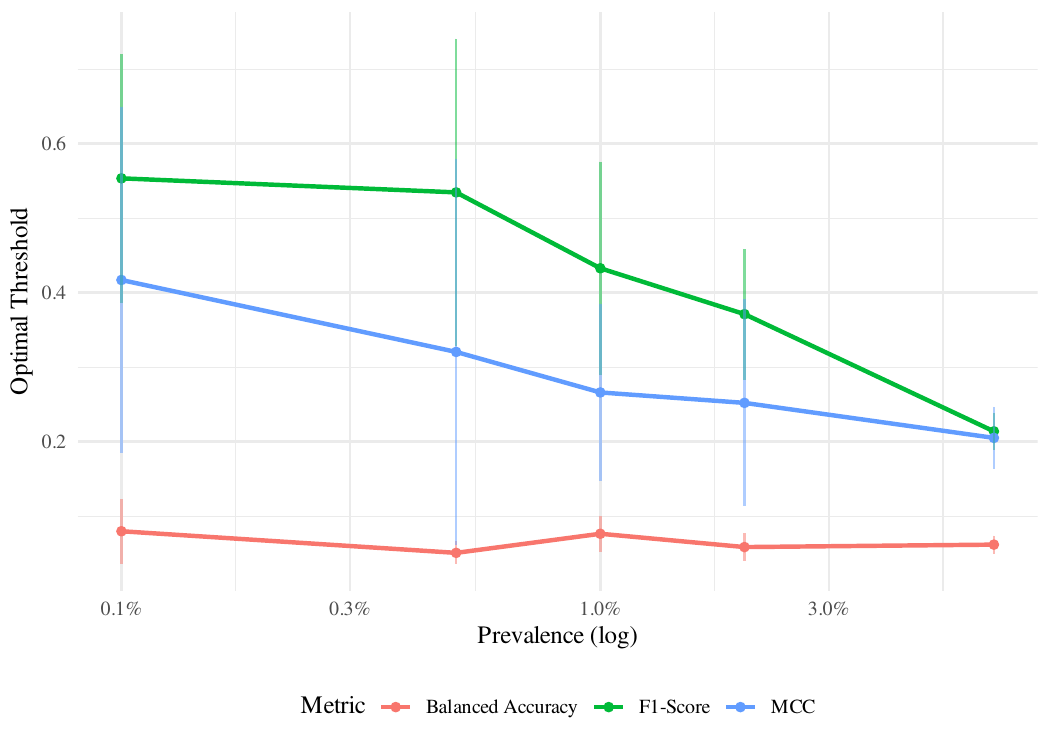}
        \caption{\textbf{Application Series D1: Classical Threshold Collapse.} Empirical threshold drift for F1, MCC, and Balanced Accuracy across prevalence regimes.}
        \label{fig:app_D1}
    \end{figure}

    \begin{figure}[H]
        \centering
        \includegraphics[width=0.75\linewidth]{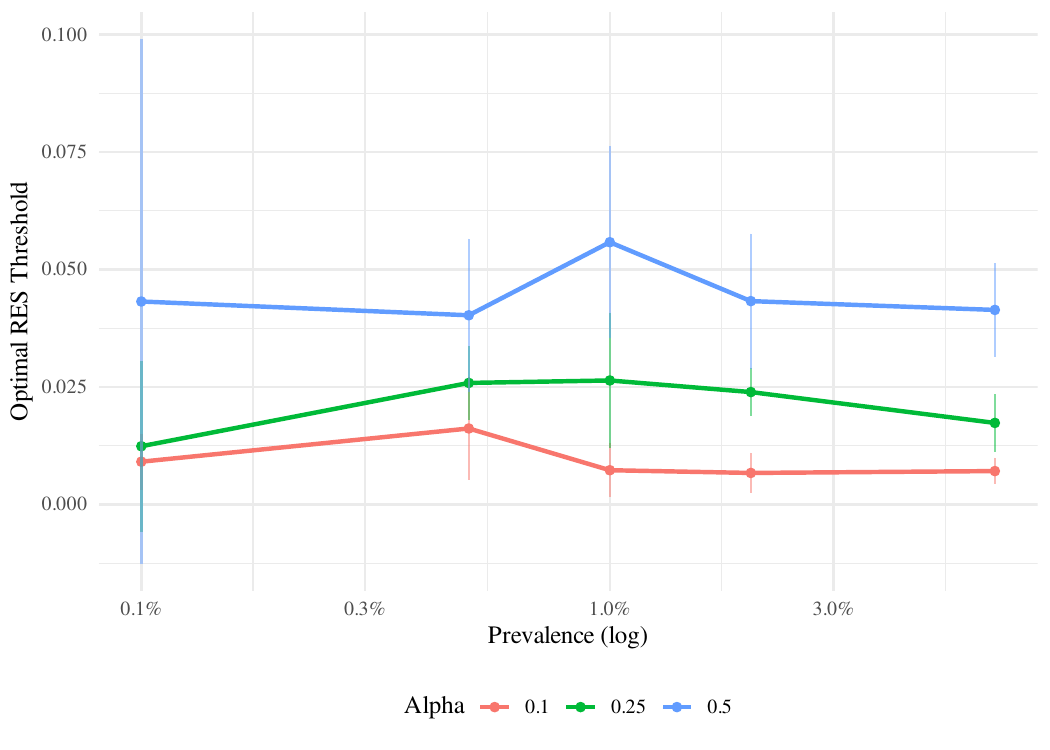}
        \caption{\textbf{Application Series D2: RES Stability.} RES threshold paths for $\alpha \in \{0.10, 0.25, 0.50\}$ across empirical prevalence regimes.}
        \label{fig:app_D2}
    \end{figure}

    \begin{figure}[H]
        \centering
        \includegraphics[width=0.75\linewidth]{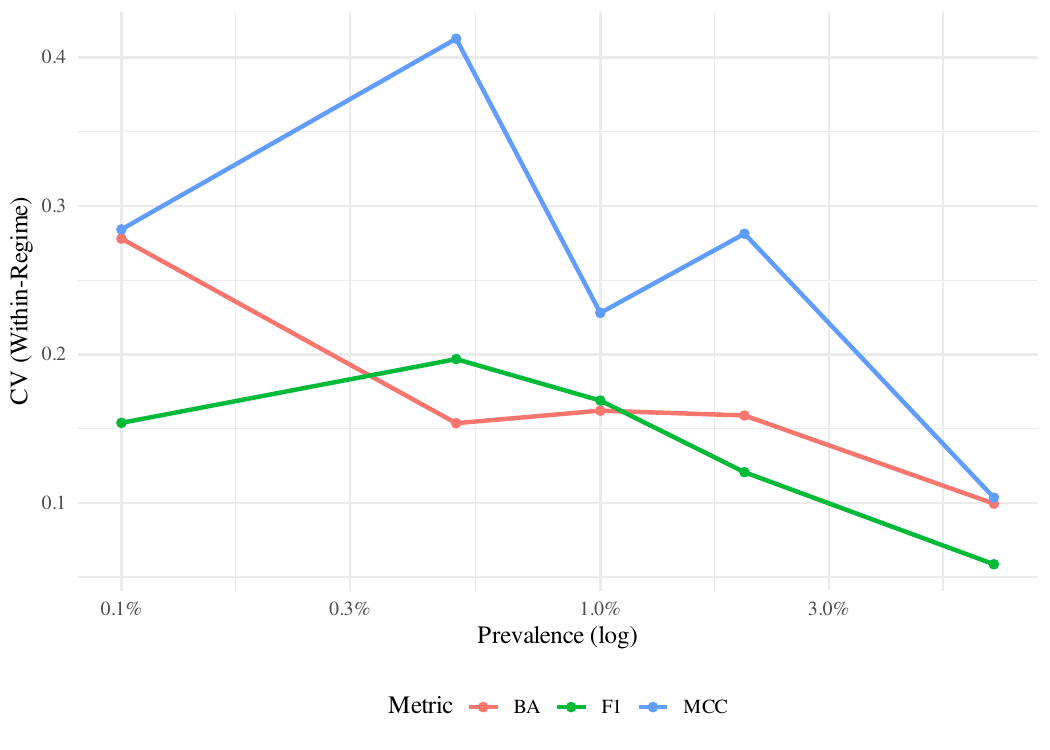}
        \caption{\textbf{Application Series D3: Within-Regime Volatility.} Coefficient of variation of classical metric thresholds within each prevalence regime.}
        \label{fig:app_D3}
    \end{figure}

    \begin{figure}[H]
        \centering
        \includegraphics[width=0.75\linewidth]{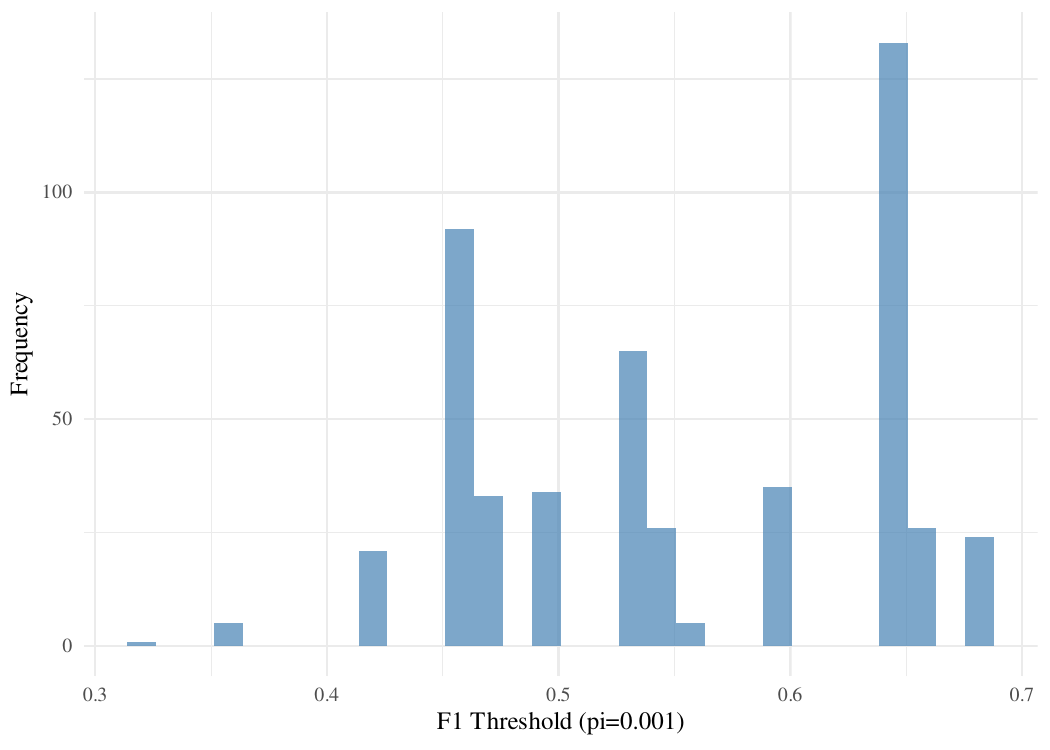}
        \caption{\textbf{Application Series F1: Rare-Regime Instability.} Bootstrap distribution of F1-optimal thresholds in the rarest regime ($\pi \approx 0.001$).}
        \label{fig:app_F1}
    \end{figure}

    \begin{figure}[H]
        \centering
        \includegraphics[width=0.75\linewidth]{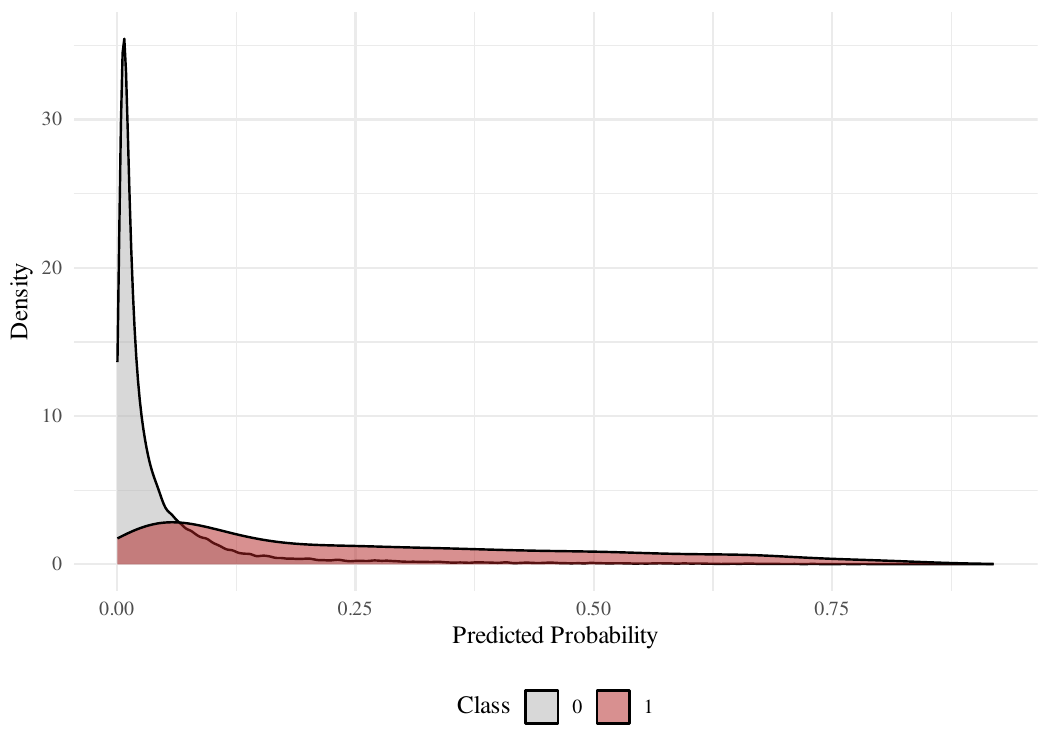}
        \caption{\textbf{Application Series D4: Conditional Score Distributions.} Density of LightGBM predicted probabilities for defaults ($Y=1$) and non-defaults ($Y=0$) in the test set. The overlap between classes confirms that the empirical setup represents a realistic, non-separable credit risk problem, justifying the need for a calibrated decision threshold.}
        \label{fig:app_D4}
    \end{figure}

    % ============================================================
    % APPENDIX I: REPRODUCIBILITY
    % ============================================================
    \newpage
    \section{Reproducibility and Code Availability}
    \label{app:reproducibility}

    All simulation code, empirical analysis scripts, and data-preprocessing routines used
    in the preparation of this article are provided in the author data and code repository
    accompanying the submission. The repository includes:

    \begin{itemize}
        \item \textbf{Simulation Scripts:}
        All routines required to reproduce the rare-event experiments of Section~6,
        including the full data-generating processes, threshold-path construction,
        metric evaluation, and summary-statistic generation.

        \item \textbf{Credit-Default Application Scripts:}
        The complete analysis pipeline for Section~7, including data cleaning, LightGBM
        probability-of-default forecasting, construction of controlled prevalence regimes,
        and bootstrap evaluation of all metrics.

        \item \textbf{Calibration Scripts:}
        The code used to implement the calibration procedures described in Appendix~E
        and applied empirically in Appendix~F.

        \item \textbf{Figure and Table Generation:}
        Scripts producing all figures and tables in the main text and Appendices.

        \item All diagnostic figures referenced in the main text and appendix are generated automatically by the provided scripts and can be reproduced end-to-end using the repository workflow.
    \end{itemize}

    % ============================================================
    % APPENDIX J: NOTATION (AT END)
    % ============================================================
    \newpage
    \section{Notation and Acronyms}
    \label{app:notation}
    \setcounter{table}{0}

    This appendix summarises the notation and the abbreviations used throughout the paper.
    All notation refers to binary classification with probabilistic forecasts
    $\eta(x) \in [0,1]$ and threshold-based alarm rules
    $\mathbf{1}\{\eta(x) \ge \delta\}$.

    \subsection{Notation}

    \begin{table}[h!]
        \centering
        \caption{Notation Summary}
        \label{tab:notation_summary}
        \begin{tabular}{ll}
            \toprule
            \textbf{Symbol} & \textbf{Meaning} \\
            \midrule
            $Y \in \{0,1\}$ & Binary outcome (1 = event). \\
            $X$ & Feature vector. \\
            $\eta(x)$ & Predicted probability or score. \\
            $\pi = \Pr(Y=1)$ & Event prevalence. \\
            $\delta$ & Alarm threshold. \\
            $\delta^{\ast}$ & Metric-induced optimal threshold. \\
            \addlinespace[3pt]
            $\TP, \FP, \TN, \FN$ & Confusion-matrix components. \\
            $\TPR(\delta)$ & True positive rate. \\
            $\FPR(\delta)$ & False positive rate. \\
            $\TNR(\delta)$ & Specificity ($1-\FPR$). \\
            $\mathrm{Prec}(\delta)$ & Precision. \\
            \addlinespace[3pt]
            $F_1(t), F_0(t)$ & Conditional CDFs of $\eta(X)$ given $Y=1$ or $0$. \\
            $f_1(t), f_0(t)$ & Conditional densities. \\
            $\Lambda(t)$ & Likelihood ratio $f_1(t)/f_0(t)$. \\
            \addlinespace[3pt]
            $\alpha$ & RES policy preference parameter. \\
            $M_{\mathrm{RE}}(\delta)$ & RES metric. \\
            $C_{\mathrm{FP}}, C_{\mathrm{FN}}$ & False-positive / false-negative costs. \\
            $\delta^{\ast}_{\text{loss}}$ & Loss-minimising threshold. \\
            \bottomrule
        \end{tabular}
    \end{table}

    \newpage

    \subsection{Acronyms}

    \begin{table}[h!]
        \centering
        \caption{Acronyms Used in the Paper}
        \label{tab:acronyms}
        \begin{tabular}{ll}
            \toprule
            \textbf{Acronym} & \textbf{Definition} \\
            \midrule
            AUC & Area Under the ROC Curve. \\
            AUPRC & Area Under the Precision--Recall Curve. \\
            BA & Balanced Accuracy. \\
            CV & Coefficient of Variation. \\
            FNR & False Negative Rate. \\
            FPR & False Positive Rate. \\
            MCC & Matthews Correlation Coefficient. \\
            PD & Probability of Default. \\
            RES & Rare-Event-Stable metrics. \\
            ROC & Receiver Operating Characteristic. \\
            TNR & True Negative Rate. \\
            TPR & True Positive Rate. \\
            \bottomrule
        \end{tabular}
    \end{table}

\end{document}